\documentclass[twocolumn,aps,superscriptaddress,pre]{revtex4}

\usepackage{amsmath,amssymb,graphicx}
\usepackage{subfigure}
\usepackage{algorithmic}
\usepackage{enumerate}
\usepackage{times}
\usepackage{color}
\usepackage{soul}
\definecolor{yblue}{rgb}{0.06, 0.3, 0.57}
\usepackage[pdftex]{hyperref}
\hypersetup{colorlinks=true,linkcolor=blue,citecolor=blue,urlcolor=blue}

\newcommand{\state}[1]{$\mathcal{S}_{#1}$}

\begin{document}

\title{Systematic solitary waves from their linear limits in two-component Bose-Einstein condensates with unequal dispersion coefficients}

\author{Wenlong Wang}
\email{wenlongcmp@scu.edu.cn}
\affiliation{College of Physics, Sichuan University, Chengdu 610065, China}

\begin{abstract}
We systematically construct vector solitary waves in harmonically trapped one-dimensional two-component Bose-Einstein condensates with unequal dispersion coefficients by a numerical continuation in chemical potentials from the respective analytic low-density linear limits to the high-density nonlinear Thomas-Fermi regime. The main feature of the linear states herein is that the component with the larger quantum number has instead a smaller linear eigenenergy, enabled by suitable unequal dispersion coefficients, leading to new series of solutions compared with the states similarly obtained in the equal dispersion setting. Particularly, the lowest-lying series gives the well-known dark-anti-dark waves, and the second series yields the dark-multi-dark states, and the following series become progressively more complex in their wave structures. The Bogoliubov-de Gennes spectra analysis shows that most of these states are typically unstable, but they can be long-lived and most of them can be fully stabilized in suitable parameter regimes.
\end{abstract}

\maketitle

\section{Introduction}
Multicomponent Bose-Einstein condensates (BECs) \cite{becbook1,becbook2} have attracted considerable attention over the past decades, providing an excellent playground for investigating a diverse array of research themes such as phase separations, spin-dependent interactions, and vector solitary waves. Vector solitary waves frequently show richer structures as well as dynamics than their one-component counterparts, e.g., they can have novel forms of instabilities \cite{Yan:DD,Lichen:DB,Lichen:RW,Wang:VRB}. In addition, these fascinating waves and the concept of a vector order parameter are also very relevant in nonlinear optics \cite{DSoptics}, the $p$-wave superfluid $^3$He, and the $d$-wave unconventional superconductors, generating very intriguing topological defects such as disclinations, cluster-forming vortices \cite{Wang:CG}, and vortices with a half-integer circulation \cite{SCbook}. 

The diversity of vector solitary waves is already evident in the one-dimensional setting. Here, we restrict our attention to the most common repulsive condensates.
The most prototypical vector soliton  in a two-component condensate is arguably the dark-bright (DB) soliton \cite{DBS1,Rajendran_2009,DDDB,DBcounterflow,Dong:MDB,DBtunneling,DBcollisions}, where a bright mass of one component is trapped and waveguided by a dark soliton of the other component. The idea that a bright mass is trapped by a density dip is rather common, e.g., there are vortex-bright and vortex-ring-bright structures as well in higher dimensions \cite{PK:VB,Wang:VRB}. However, not all vector solitons are of this type. There are stationary as well as beating dark-dark structures, and the dark-bright and beating dark-dark states are mathematically related when the system possesses the $SO(2)$ symmetry \cite{Yan:DD,Wang:SO2}. In recent years, even more intriguing structures are found, particularly the dark-anti-dark (DAD) waves \cite{engels16,engels20}, e.g., the magnetic soliton \cite{string,MS:20a,MS:20b}. Here, the anti-dark soliton is a ``bright'' soliton sitting on top of a finite rather than zero background. 
In addition, there are also states where the trapped ``bright'' component is excited showing one or even multiple phase windings, which we refer to as dark-multi-dark states herein \cite{DBunequal,VBunequal}. 



Many techniques have been developed for finding solitary wave solutions, see \cite{Wang:MDDD} for a brief discussion of some established analytic methods and numerical methods, e.g., the Darboux transformation \cite{Lichen:DT} and the deflation method \cite{Panos:DC3}, respectively. The linear limit continuation method \cite{Wang:DD,Wang:MDDD} is a hybrid numerical continuation technique, it starts from an analytically tractable linear limit, effectively turning off the nonlinearity, and then the nonlinearity is gradually restored by slowly increasing the chemical potentials or densities. In this process, a linear wavefunction gradually deforms into a solitary wave configuration in the Thomas-Fermi (TF) regime. We briefly comment that this computational strategy is similar in spirit to the sequential Monte Carlo method in statistical physics \cite{SA,Wang:PA}.

The linear limit continuation provides a powerful and effective approach to \textit{systematically} construct solitary wave solutions from their underlying linear limits. Importantly, it also provides a framework to organize the rather diverse solitary waves. In fact, it provides an intuitive understanding on the very existence of these waves in the first place, particularly in (the nonintegrable) higher dimensions. On the numerical side, it enables a controlled approach of finding numerically exact solutions if a suitable linear limit can be identified for a solitary wave.
This method has been applied to many particular states, and recently it was systematically applied to the one-dimensional system upto five components, finding a large array of solitary waves of increasing complexity \cite{Wang:DD,Wang:MDDD}. In the two-component setting, the low-lying states are the well-known single dark-bright soliton, the in-phase two dark-bright solitons, the out-of-phase two dark-bright solitons, and so on. While many solitary waves have been found, these states are essentially of dark-bright and dark-dark structures, and therefore, they are not complete.
Here, we are interested in whether the more exotic dark-anti-dark and dark-multi-dark states can be systematically organized into this theoretical framework. From this perspective, these states have a dominant component stemming from a smaller quantum number, which couples a component stemming from a larger quantum number, exactly the opposite of the regular continuation in \cite{Wang:DD}. Our work suggests that this is possible. In addition, the setup can be readily generalized to find more complex wave patterns in a systematic manner.

The main purpose of this work is to systematically construct vector solitary waves of two components by the linear limit continuation method, where the component stemming from a linear state of smaller quantum number has instead a larger linear eigenenergy by suitably tuning the dispersion coefficients. The motivation of utilizing different dispersion coefficients was technical in origin, which we shall explain in the next section. The main function thereof is to engineer the linear eigenenergies and meanwhile also the spatial length scales of the soliton structures.
Our continuation is successful, the lowest-lying series \state{0n}, $n>0$ yields the dark-anti-dark waves, and the next series \state{1n}, $n>1$ gives the dark-multi-dark waves, and so on.


Finally, the setting of different dispersion coefficients can be experimentally implemented.
Such systems can be naturally realized in Bose mixtures of different atomic species \cite{SpinorBECs}. They are also relevant in spin-orbit coupled BECs where the coefficients arise from the curvature of the dispersion relations in different branches, and these coefficients can also be engineered \cite{SOC1,SOC2}.
Therefore, it should be possible to implement and study the solitary waves herein experimentally.
It is possible to controllably produce dark solitary waves of arbitrary speed \cite{Fritsch:DS} in a single component condensate using the density and phase engineering \cite{QE1,QE2,Fritsch:DS}. Vector solitary waves can be formed using a spatially dependent spin interconversion and density and phase imprinting using a steerable laser beam \cite{DBS2,Panos:BEC3C}.

This work is organized as follows. In Sec.~\ref{setup}, we introduce 
the model, the numerical setup, and the numerical methods. Next, we present our results in Sec.~\ref{results}. 
Finally, our conclusions and a number of open
problems for future consideration are given in Sec.~\ref{conclusion}.

\section{Model and methods}
\label{setup}

\subsection{Computational setup}

In the framework of the mean-field theory, the dynamics of one-dimensional $n$-component repulsive BECs, 
confined in a time-independent trap $V$, is described by the following coupled
dimensionless Gross-Pitaevskii equation \cite{Panos:book,Wang:DD}:
\begin{eqnarray}
i \frac{\partial \psi_j}{\partial t} = -\frac{D_j}{2} \psi_{jxx}+V \psi_j + \left(\sum_{k=1}^n g_{jk}| \psi_k |^2\right) \psi_j,
\label{GPE}
\end{eqnarray}
where $\psi_j, j=1, 2, ..., n$ are $n$ complex scalar macroscopic wavefunctions and $D_j=\kappa_j^2$ is the dispersion coefficient of the $j$th component. Here, we work with $n=2$ but keep our notations more generic as the setup is not limited to $n=2$. We set $\kappa_1=1$ without loss of generality. We focus here on the Manakov system \cite{Manakov74} of equal repulsive interactions $g_{ij}=1$ for simplicity unless otherwise specified, but the numerical setup is not limited to this constraint and further continuation in $g_{ij}$ is also possible. We ignore the spin-dependent interactions, as they are typically much smaller and can also be suppressed \cite{Panos:BEC3C}. The condensates are confined in a harmonic trap of the form: 
\begin{equation}
V=\frac{1}{2} x^2,
\label{potential}
\end{equation}
where the trapping frequency is set to $1$ by scaling without loss of generality \cite{Wang:DD}. Stationary states of the form:
\begin{eqnarray}
\psi_j(x,t) = \psi^0_j(x)e^{-i\mu_jt}
\label{ss}
\end{eqnarray}
lead to $n$ coupled stationary equations:
\begin{eqnarray}
\label{SS1}
-\frac{D_j}{2} \psi^0_{jxx}+V \psi^0_j +\left(\sum_{k=1}^n g_{jk}| \psi_k^0 |^2\right) \psi^0_j = \mu_j \psi^0_j,
\end{eqnarray}
where $\mu_j$ is the chemical potential of the $j$th component. The dimensionless scaling analysis can be found in \cite{Wang:DD,Wang:MDDD}. In summary, our length is measured in units of $\ell_x$, time in units of 
$1/\omega_x$, and the chemical potential in 
units of $\hbar \omega_x$. The number of particles in the $j$th component 
is $N_j = \frac{1}{g_{1D}} \int |\psi_j|^2 dx$, where
$g_{1D}=\frac{2\omega_{\perp}a_s}{\omega_x \ell_x}$. Here, $a_s, \omega_x, \omega_{\perp}, \ell_x=(\hbar/(m\omega_x))^{1/2}$ stand for the s-wave scatting length, the axial, and transverse trapping frequencies, and the axial harmonic oscillator length in physical units, respectively. 
It is worth mentioning that while these states are constructed in the harmonic potential, further continuation to other potentials by interpolation is accessible, showing the flexibility of the method.

Our numerical simulation for each solitary wave includes identifying a series of stationary states following a continuation path in the chemical potential parameter space, computing the Bogoliubov-de Gennes (BdG) stability spectrum along the path, and finally conducting dynamics of the solitary waves \cite{Wang:DD,Wang:MDDD}. A stationary state is found using the finite element method, and the Newton's method for convergence given a suitable initial guess; see the next section for details. For each stationary state, we compute the lowest $200$ BdG eigenvalues in magnitude. If an eigenvalue $\lambda$ has a finite real part Re$(\lambda)>0$, it signals a dynamical instability at the corresponding parameters. For states \state{35} and \state{25} with more unstable modes, we collect $300$ and $400$ eigenvalues, respectively. Therefore, the number can depend on the specific states, an insufficient signature is that the real part of some eigenvalues can suddenly terminate in the air, i.e., at finite values when $\mu_1$ is tuned. Finally, the dynamics is conducted using the regular fourth-order Runge-Kutta method.

\subsection{Linear limit continuation}
\label{ll}

In the linear limit, the fields are faint such that they decouple, and the linear limits are well-known for the harmonic trap. For a two-component system, a stationary state has two quantum numbers from the two independent harmonic oscillator states $|n_1, n_2\rangle$ \cite{Wang:OD,Wang:DD}. A linear state and its linear eigenvalues are used for finding a stationary state slightly away from the linear limit, where the chemical potentials are slightly larger than the linear eigenenergies. Then, the new solution is used for finding a stationary state at yet slightly larger chemical potentials, and so on. As the chemical potentials increase, the stationary state gradually evolves from a linear state $|n_1, n_2\rangle$ into a solitary wave \state{n_1n_2} in the TF regime. As discussed earlier \cite{Wang:DD,Wang:MDDD}, it is sufficient to focus on irreducible states with distinct quantum numbers.

When $D_j=1$, the state \state{n_1n_2} has its linear limit at the vector chemical potential $(\mu_{10}, \mu_{20})=(n_1+1/2, n_2+1/2)$. The states of $n_1>n_2\geq 0$ have already been explored \cite{Wang:DD}, finding an array of well-known states like the single DB soliton \state{10}, the in-phase two DB solitons \state{20}, and the out-of-phase two DB solitons \state{21}, and also more complex states. Here, we reverse the order and focus on the case $n_2>n_1 \geq 0$ as mentioned earlier. For example, the low-lying states are \state{01}, \state{02}, \state{12}, and so on.

It is important to explain the motivation of the continuation in the setting of unequal dispersion coefficients. In the normal order $n_1>n_2$, the linear chemical potentials have the same order $\mu_{10}>\mu_{20}$ in the regular setting of equal dispersion, equal potential function, and Manakov interactions. It is possible to increase both chemical potentials into the TF regime while maintaining this order. However, in the case $n_1<n_2$, one has $\mu_{10}<\mu_{20}$ in the regular setting. Since we want the first component to be asymptotically a dominant state in density, we require $\mu_1 \gtrsim \mu_2$ instead in the TF regime. Otherwise, we would be merely swapping the two field labels with no genuinely new state. This suggests that the continuation path should cross the line $\mu_1=\mu_2$ in the chemical potential space. To our knowledge, this is impossible in the regular setting, as the $\mu_1=\mu_2$ line is asymptotically an existence boundary. Indeed, our naive numerical continuation trying to cross this line all failed, suggesting that a new setup is necessary. This is our motivation to tune the dispersion coefficients such that the linear chemical potentials $\mu_{10}>\mu_{20}$ despite that the quantum numbers have the opposite order $n_1<n_2$. In this way, we can increase both chemical potentials while maintaining $\mu_1>\mu_2$ throughout the continuation. Our numerical work confirms that this solution is effective.

The linear stationary state equation with $\kappa_i \neq 1$ can be readily solved by scaling, one can divide the equation by $\kappa_i$ and set $y=x/\sqrt{\kappa_i}$. After some straightforward algebra, we find that the linear state $|m\rangle$ has the eigenenergy $(m+1/2)\kappa_i$, and the state is $\phi_m(x/\sqrt{\kappa_i})$ where $\phi_m(x)$ is the regular quantum harmonic oscillator state of $\kappa=1$. This means that if $\kappa_i$ gets smaller, then its eigenenergy gets smaller and the wavefunction simultaneously takes a smaller spatial profile. These linear state solutions are also confirmed numerically using the finite element method by directly diagonalizing the Hamiltonian matrix in the position representation. In this work, we set $\kappa_1=1$ and $\kappa_2=\kappa$ unless otherwise specified.

To satisfy $\mu_{10}>\mu_{20}$ for state \state{n_1n_2} at the linear limit, we need that $(n_1+1/2)>(n_2+1/2)\kappa$ or $\kappa<(2n_1+1)/(2n_2+1)$. In this work, we choose for simplicity 
\begin{align}
\kappa = \frac{2n_1+1}{2n_2+1.2},
\label{kappa}
\end{align}
unless specified otherwise. It is noted that the number $1.2$ is not special in any particular way, it is only chosen such that $\kappa$ is not unnecessarily small, as previous works suggest that this helps towards the stabilization of such states \cite{DBunequal,VBunequal}.

\begin{table*}[htb]
\caption{
Simulation parameters and summary of stable intervals for the solitary waves studied in this work, see Eqs.~(\ref{kappa}-\ref{trajectory}) for details. The last column means that the stability is dynamically confirmed for the selected state upto $t=1000$.
\label{para1}
}
\begin{tabular*}{\textwidth}{@{\extracolsep{\fill}} l c c c c r}
\hline
\hline
States &$\Delta_f$ &$\alpha$ &$\mu_{1\mathrm{max}}$ &Stability &Dynamics \\
\hline
\state{01} &$\Delta_0/5$ &$0.5$ &$16$ &Fully stable &$\mu_1=16.0$ \\
\state{02} &$\Delta_0/5$ &$0.5$ &$16$ &Stable $\mu_1 \in (0.5, 3.41] \cup[3.97, 6.86]$ &$\mu_1=6.8$ \\
\state{03} &$\Delta_0/5$ &$0.5$ &$16$ &Stable $\mu_1 \in (0.5, 3.37]$ &$\mu_1=3.3$ \\
\state{04} &$\Delta_0/5$ &$0.5$ &$16$ &Stable $\mu_1 \in (0.5, 1.59] \cup [1.74, 2.68]$ &$\mu_1=2.5$ \\
\state{05} &$\Delta_0/5$ &$0.5$ &$16$ &Stable $\mu_1 \in [1.43, 2.10] \cup [2.47, 3.05]$ &$\mu_1=3.0$ \\
\state{06} &$\Delta_0/5$ &$0.5$ &$16$ &Stable $\mu_1 \in [1.28, 1.57] \cup [1.97, 3.02]$ &$\mu_1=2.9$ \\
\state{07} &$\Delta_0/5$ &$0.5$ &$16$ &Stable $\mu_1 \in (0.5, 1.38] \cup [1.65, 1.94]$ &$\mu_1=1.9$ \\
\state{08} &$\Delta_0/5$ &$0.5$ &$16$ &Stable $\mu_1 \in (0.5, 1.27] \cup [2.10, 3.12]$ &$\mu_1=3.0$ \\
\state{09} &$\Delta_0/5$ &$0.5$ &$16$ &Stable $\mu_1 \in (0.5, 1.33] \cup [1.78, 1.97]$ &$\mu_1=1.9$ \\
\state{010} &$\Delta_0/5$ &$0.5$ &$16$ &Stable $\mu_1 \in (0.5, 1.38] \cup [1.84, 1.97] \cup [2.20, 2.51]$, etc. &$\mu_1=2.4$ \\
\state{12} &$\Delta_0/3$ &$0.5$ &$16$ &Stable $\mu_1 \in [4.19, 4.99] \cup [6.19, 6.83] \cup [12.90, 13.69]$, etc. &$\mu_1=13.5$ \\
\state{13} &$\Delta_0/3$ &$0.5$ &$16$ &Stable $\mu_1 \in [6.17, 6.35] \cup [10.74, 11.02] \cup [13.77, 14.24]$, etc. &$\mu_1=14.0$ \\
\state{14} &$\Delta_0/3$ &$0.5$ &$45$ &Stable $\mu_1 \in [28.79, 28.91] \cup [36.48, 36.81] \cup [43.73, 43.82]$, etc. &$\mu_1=36.6$ \\
\state{15} &$\Delta_0/3$ &$0.5$ &$50$ &Stable $\mu_1 \in [35.43, 35.79] \cup [41.13, 41.69] \cup [45.88, 47.22]$, etc. &$\mu_1=46.0$ \\
\state{23} &$\Delta_0/3$ &$0.5$ &$16$ &Stable $\mu_1 \in [4.67, 4.90] \cup [6.43, 6.72] \cup [15.85, 15.92]$, etc. &$\mu_1=15.9$ \\
\state{24} &$\Delta_0/3$ &$0.5$ &$20$ &Stable $\mu_1 \in [5.61, 5.64]$ &$\mu_1=5.63$ \\
\state{25} &$\Delta_0/3$ &$0.5$ &$50$ &Unstable &- \\
\state{34} &$\Delta_0/3$ &$0.5$ &$20$ &Stable $\mu_1 \in [5.81, 6.28] \cup [8.30, 8.64] \cup [8.87, 9.06]$, etc. &$\mu_1=9.0$ \\
\state{35} &$\Delta_0/3$ &$0.5$ &$50$ &Unstable &- \\
\state{45} &$\Delta_0/3$ &$0.5$ &$30$ &Stable $\mu_1 \in [7.51, 7.88] \cup [10.53, 10.84] \cup [11.53, 11.70]$, etc. &$\mu_1=11.6$ \\
\state{89} &$\Delta_0/2.5$ &$0.5$ &$30$ &Stable $\mu_1 \in [24.47, 24.60]$ &$\mu_1=24.5$ \\
\state{910} &$\Delta_0/3$ &$0.5$ &$30$ &Stable $\mu_1 \in [26.56, 26.91]$ &$\mu_1=26.8$ \\
\hline
\hline
\end{tabular*}
\end{table*} 

Finally, we present the continuation details in the chemical potential parameter space. In the earlier work \cite{Wang:DD}, we utilized a simple linear trajectory. As the states herein are more intriguing, we follow a more flexible exponential path:
\begin{align}
    \mu_{2} = \mu_1- \Delta_{f} + (\Delta_{f}-\Delta_0)\exp(-\alpha(\mu_1-\mu_{10})),
\label{trajectory}
\end{align}
where the initial gap $\Delta_0=\mu_{10}-\mu_{20}$. The asymptotic gap is $\Delta_f$ and $\alpha$ controls how rapidly the path converges to the asymptotic form. In our numerical work, $\mu_1$ runs from the linear limit $\mu_{10}$ to $\mu_{1\mathrm{max}}$ in the TF regime. The simulation parameters are summarized in Table~\ref{para1}.
When the chemical potentials are reasonably large, the existence of a state becomes pretty robust such that one can further continue it in other parameters, e.g., $\mu_2$ and $g_{ij}$, in line with many previous works \cite{Wang:DD,Wang:MDDD,Wang:VRB}.




\section{Numerical results}
\label{results}

The numerical continuation is successful, yielding arrays of solitary wave solutions from their linear limits. As mentioned earlier, the lowest-lying $S_{0n}$ states are the dark-anti-dark waves, and the second series of $S_{1n}$ states are the dark-multi-dark waves, and there are solitary wave patterns of increasing complexity from the $S_{2n}$ states, $S_{3n}$ states, and so on. This large set of solitary waves is ideal for further exploring their properties. To this end, we illustrate two show-of-principle unstable scenarios.

\subsection{Dark-anti-dark waves}

\begin{figure*}[htb]
\subfigure[]{\includegraphics[width=0.195\textwidth]{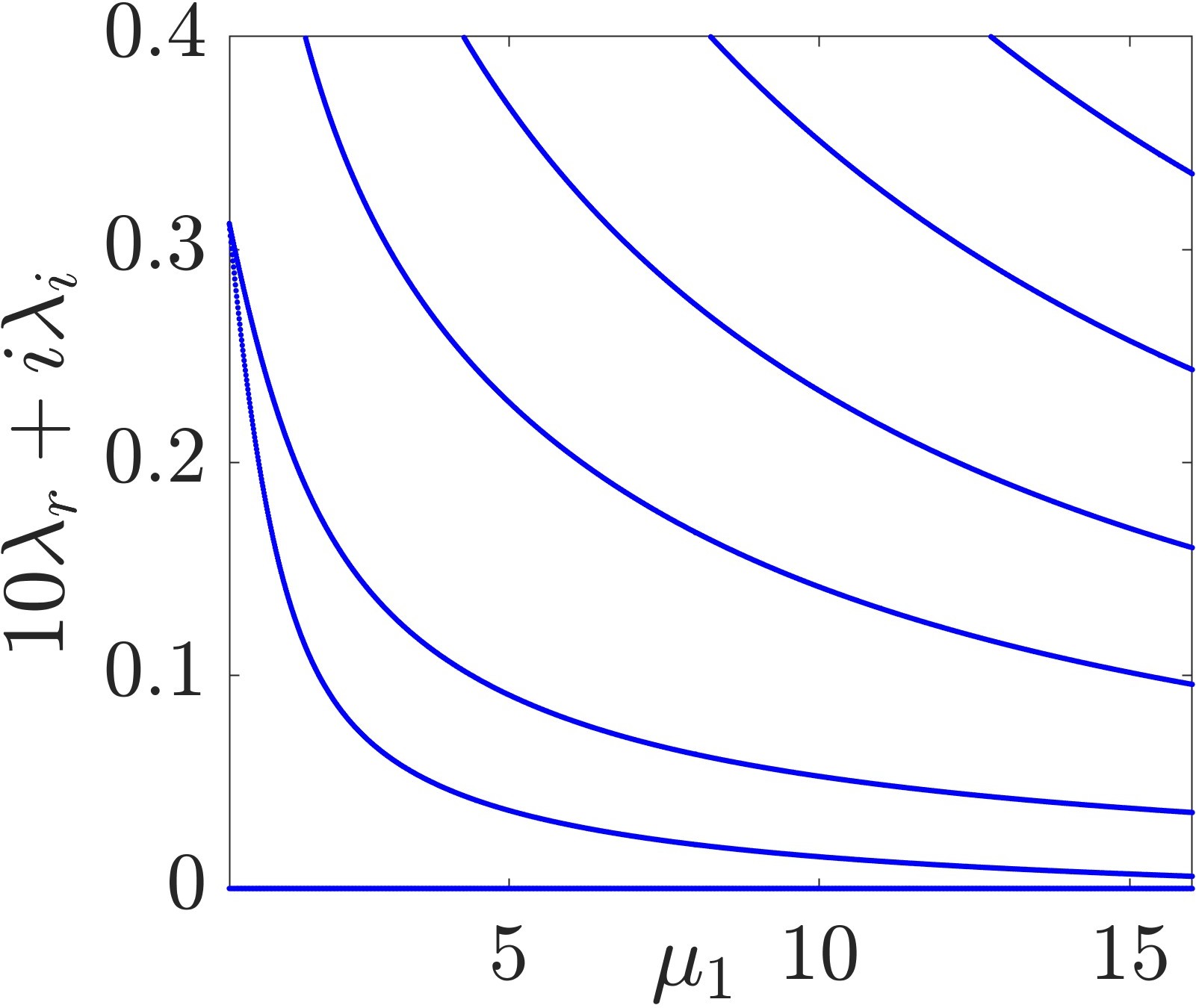}}
\subfigure[]{\includegraphics[width=0.195\textwidth]{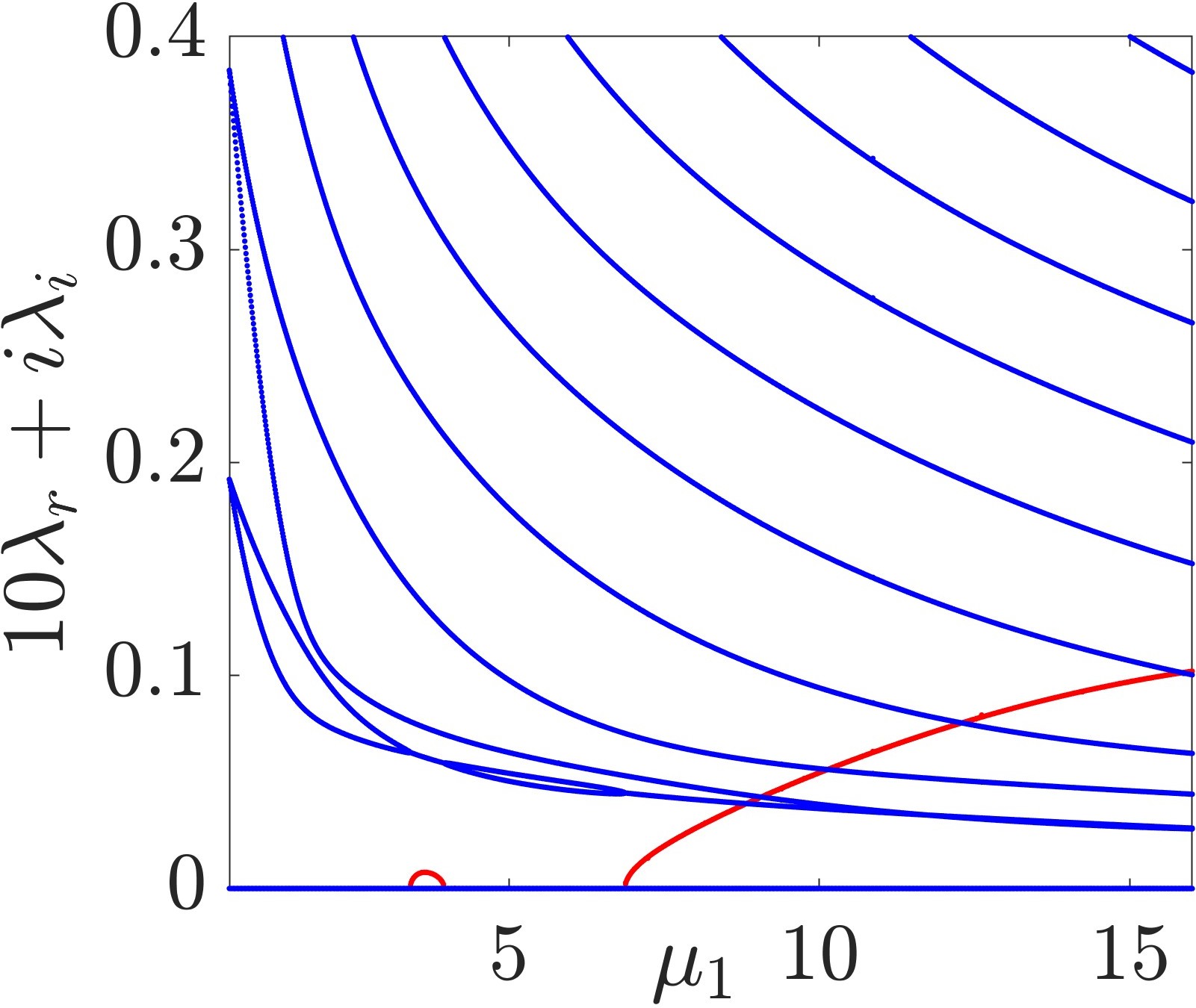}}
\subfigure[]{\includegraphics[width=0.195\textwidth]{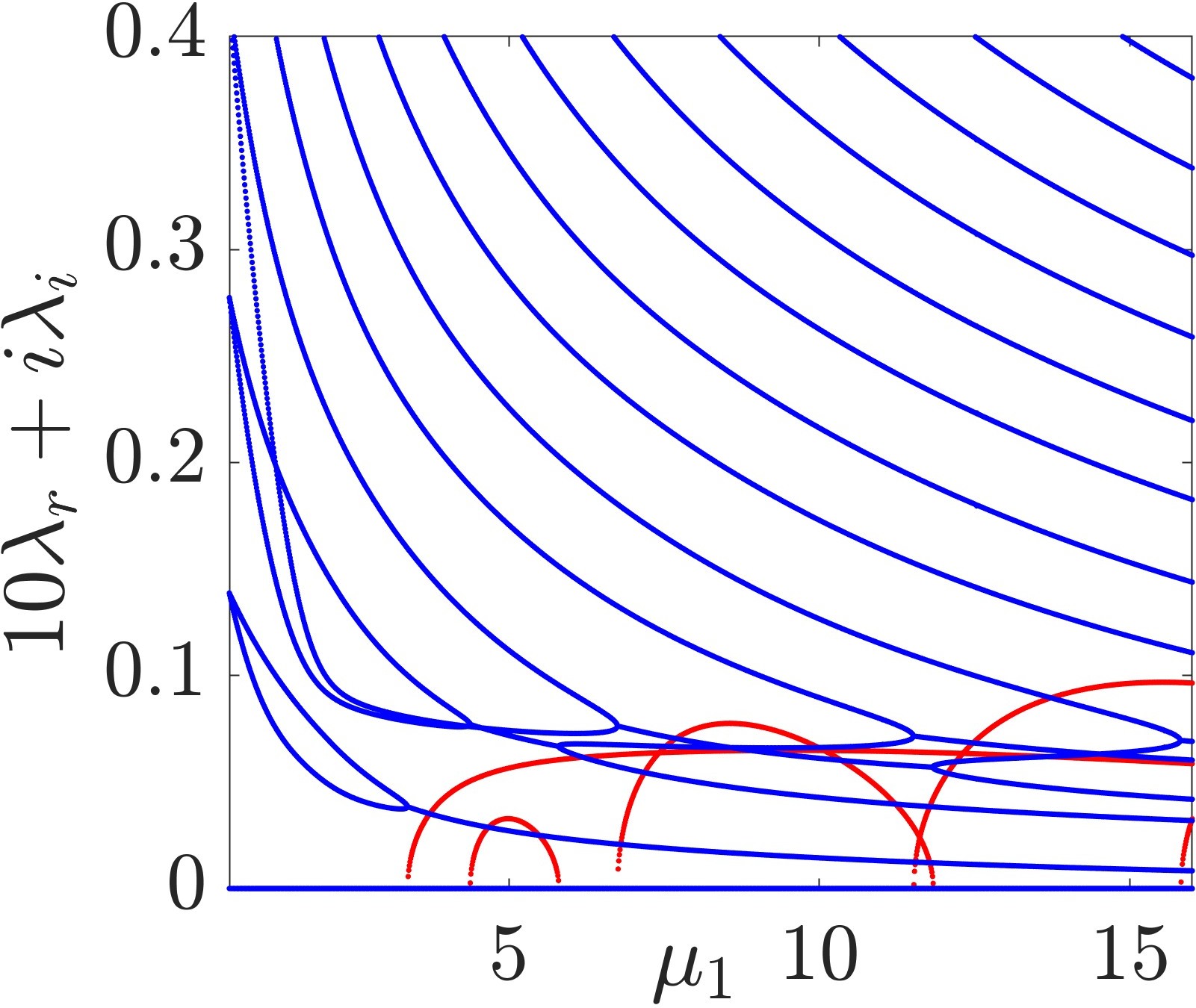}}
\subfigure[]{\includegraphics[width=0.195\textwidth]{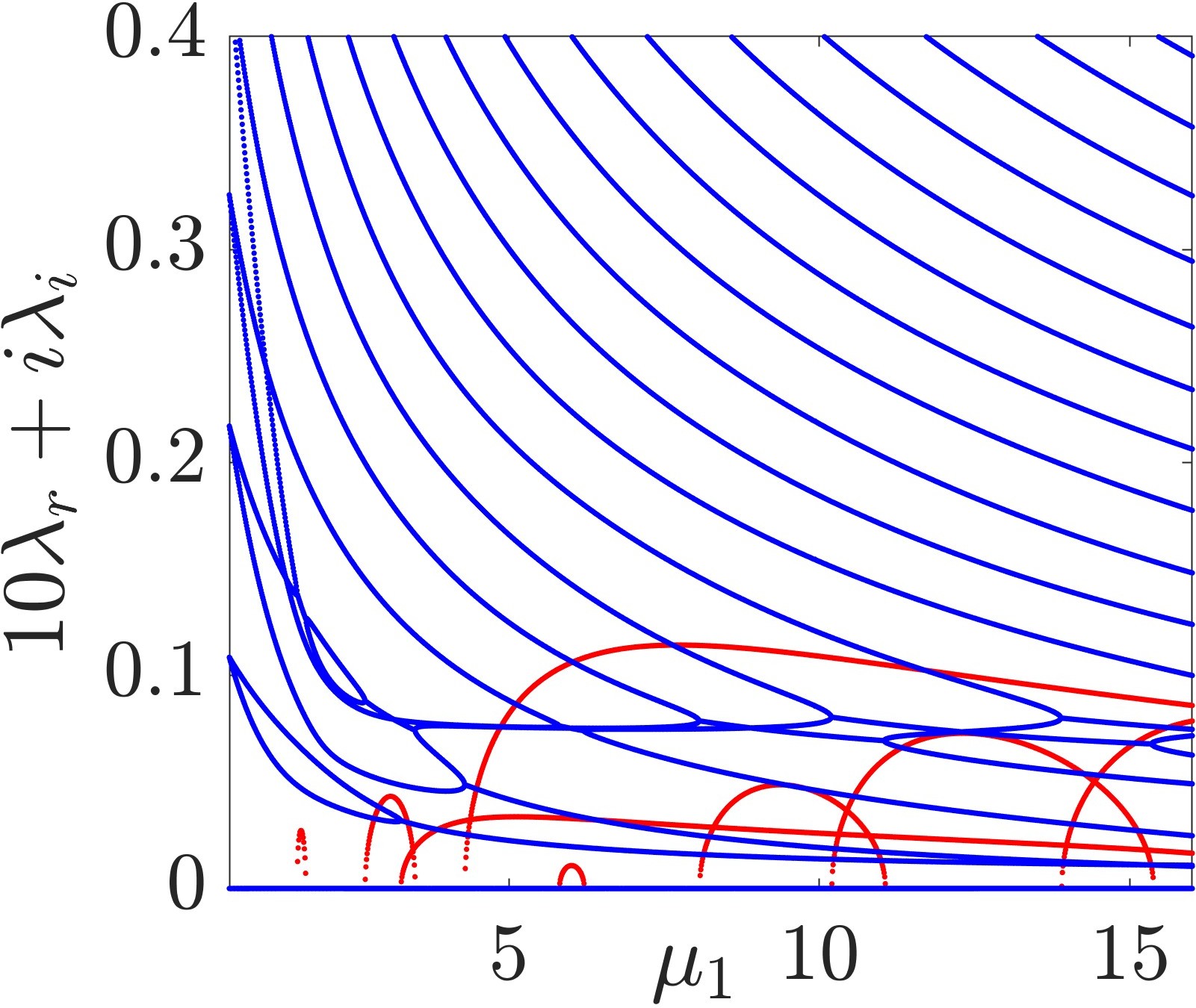}}
\subfigure[]{\includegraphics[width=0.195\textwidth]{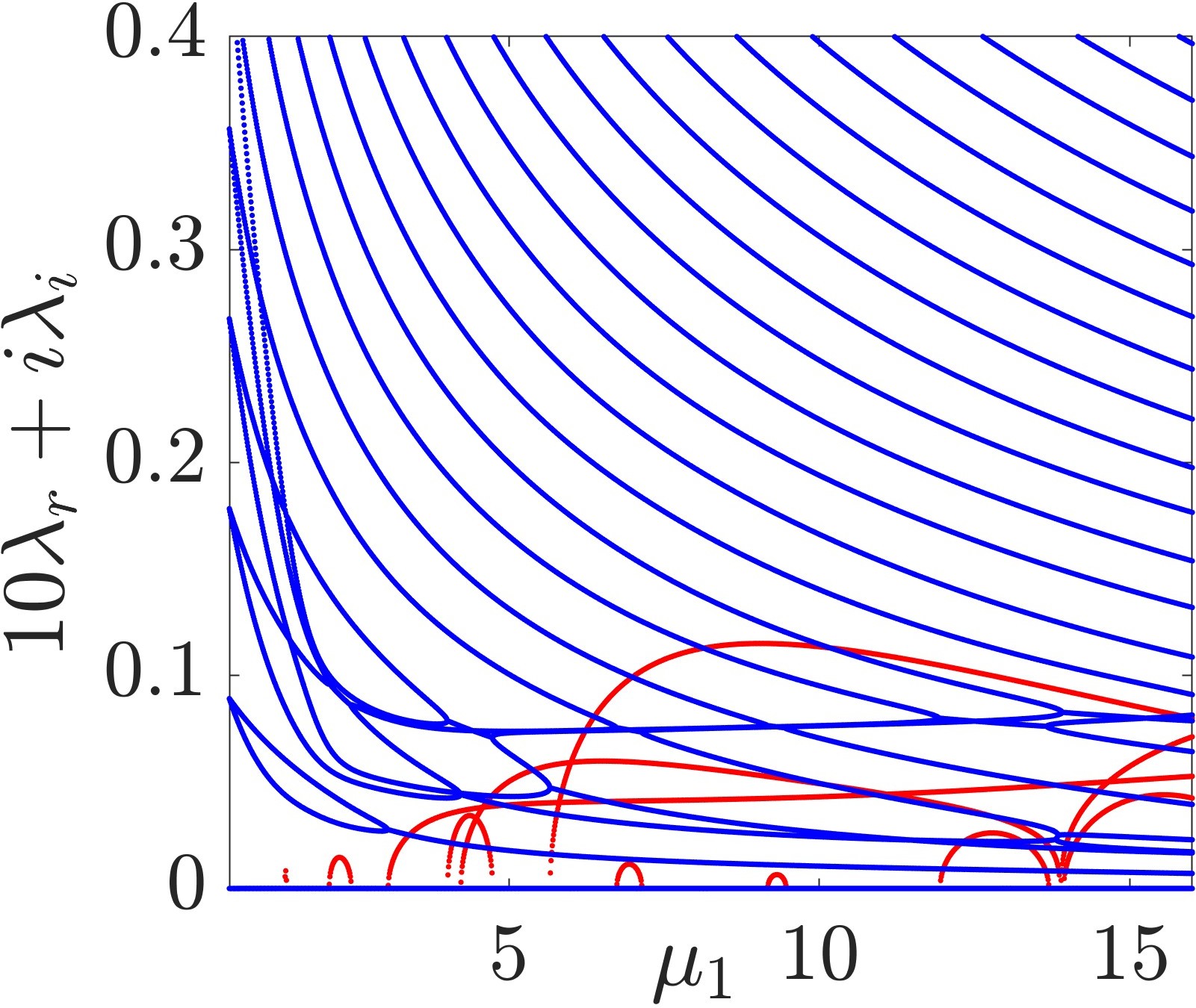}}
\subfigure[]{\includegraphics[width=0.195\textwidth]{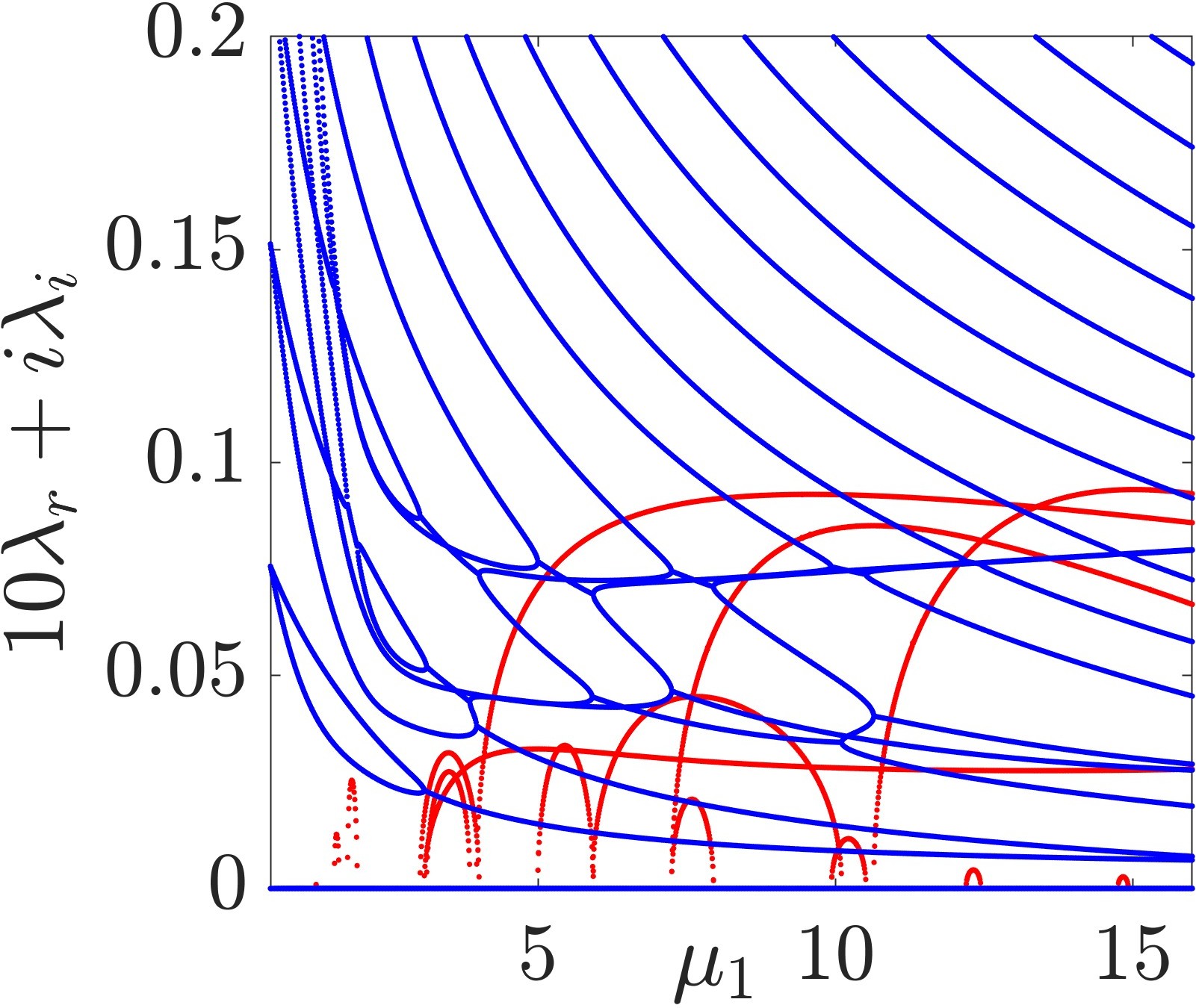}}
\subfigure[]{\includegraphics[width=0.195\textwidth]{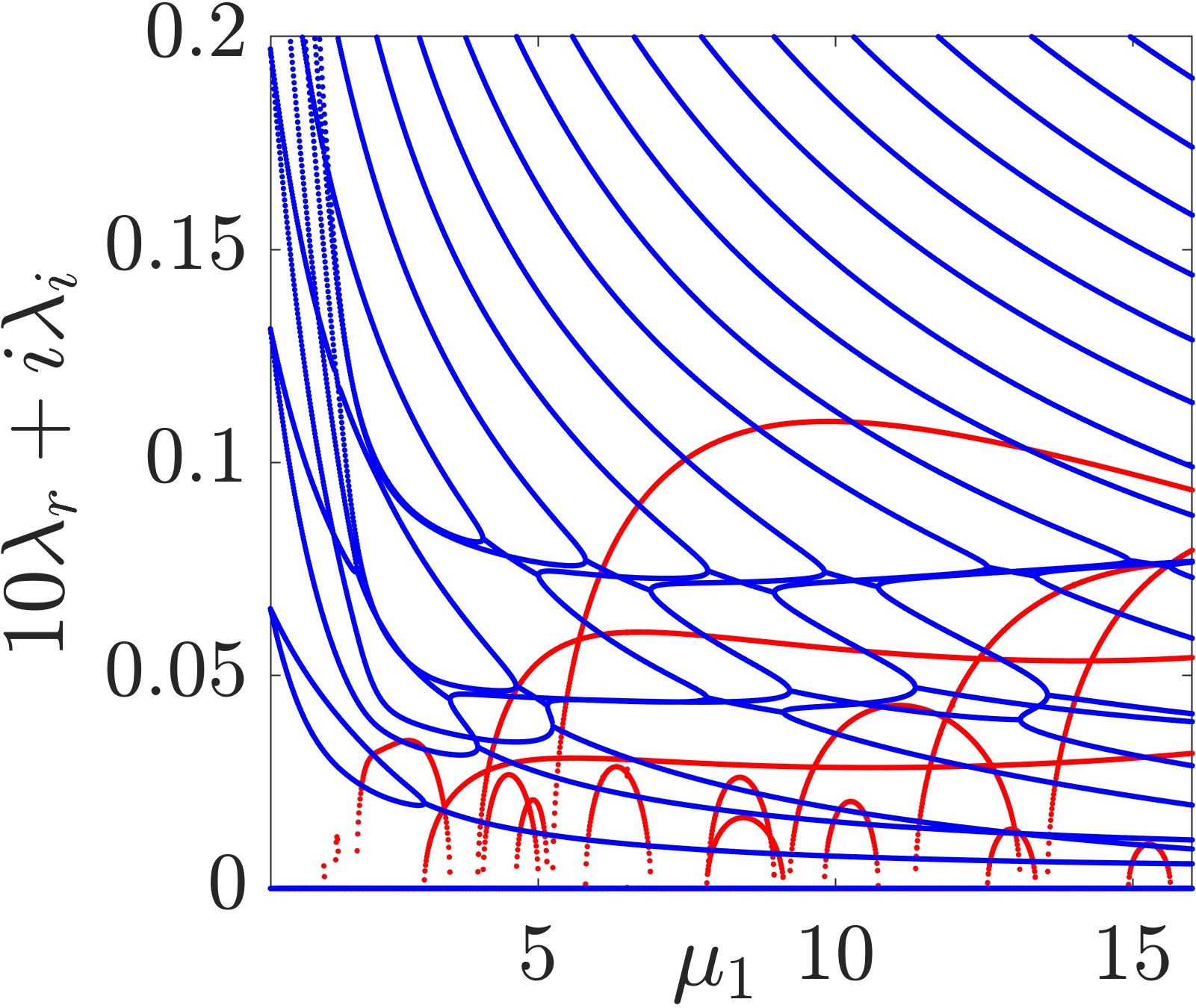}}
\subfigure[]{\includegraphics[width=0.195\textwidth]{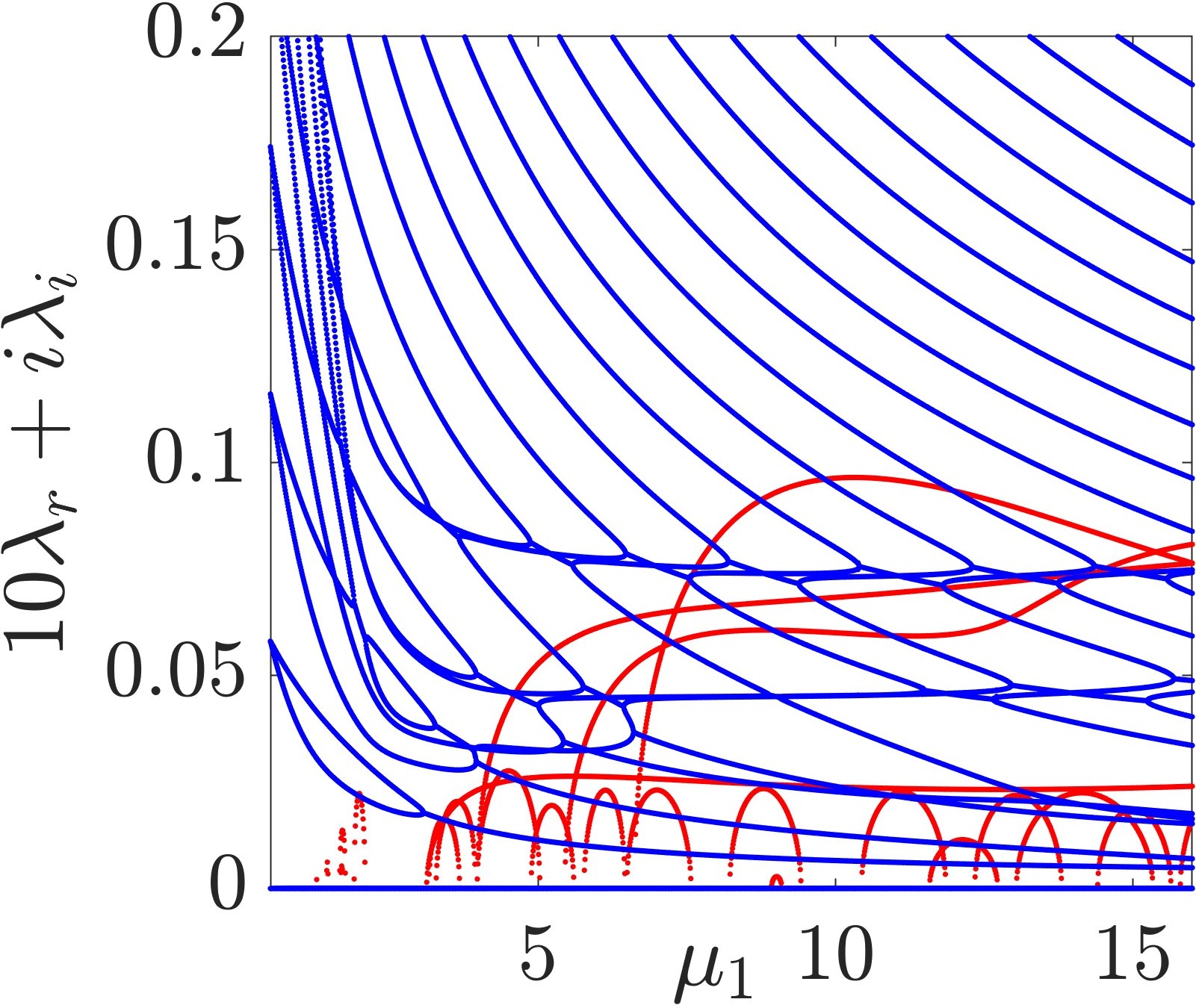}}
\subfigure[]{\includegraphics[width=0.195\textwidth]{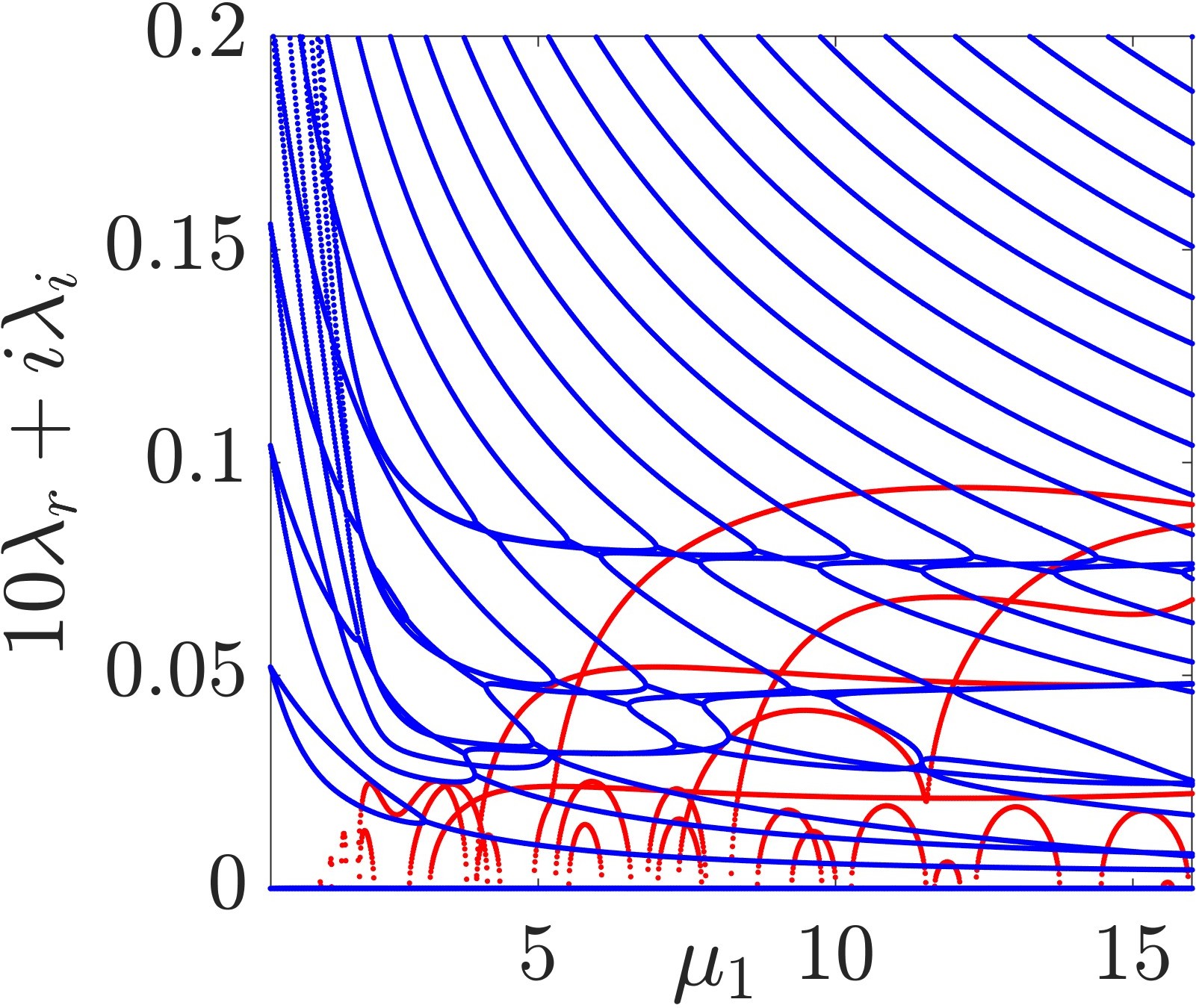}}
\subfigure[]{\includegraphics[width=0.195\textwidth]{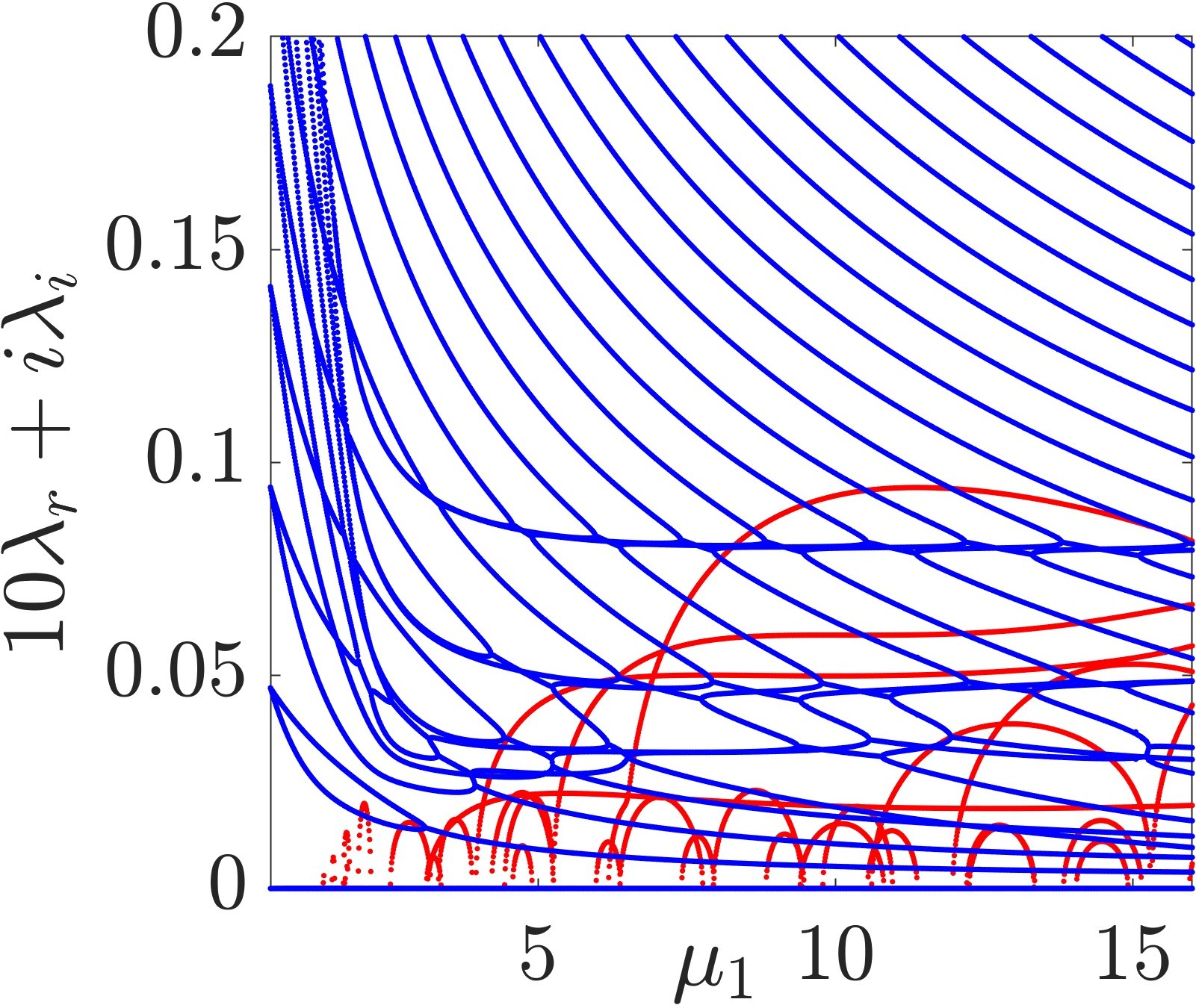}}
\subfigure[]{\includegraphics[width=\textwidth]{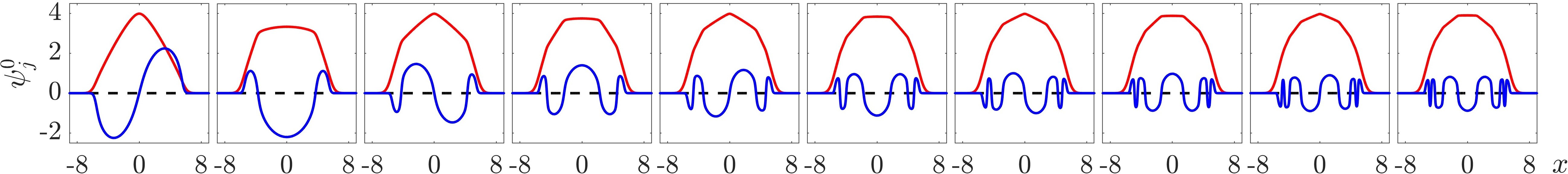}}
\subfigure[]{\includegraphics[width=\textwidth]{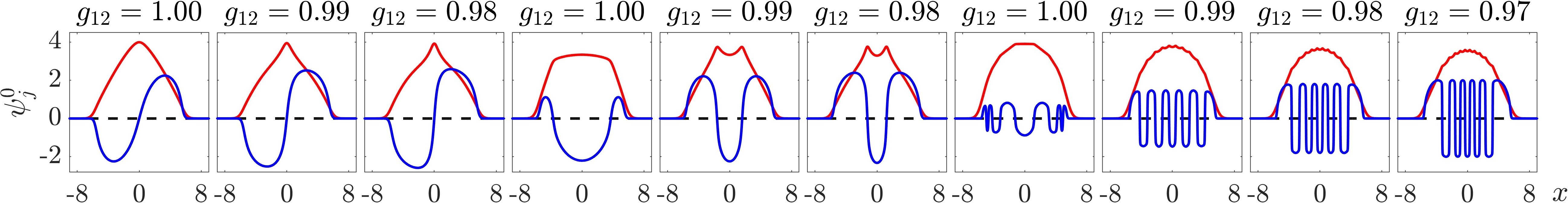}}
\caption{
The BdG spectra (a-j) $\lambda$ of states \state{01}, \state{02}, \state{03}, and so on up to \state{010}, respectively, along the respective continuation trajectories from the linear limits to a typical large-density TF regime, see also Table~\ref{para1}. Red (the low-lying modes stemming from zero) and blue lines are for the (unstable) real and (stable) imaginary parts of the eigenvalues, respectively. As the instability is typically very weak, the real part of the eigenvalues is enlarged by a factor of $10$ for clarity. Panel (k) shows some typical configurations, $\psi_1^0$ in red and $\psi_2^0$ in blue, at $\mu_1=16$ for the respective states above. Panel (l) shows that the states take more regular dark-anti-dark waveforms as the intercomponent interaction $g_{12}$ is lowered, here, we only illustrate a few typical states of \state{01}, \state{02}, and \state{010} for simplicity, but this holds for all the states above.
}
\label{S01}
\end{figure*}

We start from the \state{0n} states, i.e., the first component is in the uniform phase ground state, while the second component contains $n=1, 2, 3, ...$ dark solitons. These states and their BdG spectra are depicted in Fig.~\ref{S01}. First, all of these states exist, suggesting that the continuation method works. The lowest-lying state \state{01} in this setup appears to be fully robust, i.e., the BdG eigenvalues are purely imaginary with no real part. This is reasonable, as both the ground state and the single dark soliton state are robust in the one-component setting, similar to the robust single dark-bright soliton \cite{Wang:DD}. The rest of the states suffer from instabilities, and the number of unstable modes appears to grow as $n$ increases, as the states bear a growing number of the so-called negative energy
modes \cite{Panos:book}. Indeed, the instabilities herein are oscillatory ones, and each unstable mode has a complex quartet of engenvalues via
the Hamiltonian-Hopf bifurcation. However, it is worth noting that the growth rates are typically quite weak $O(10^{-2})$; cf. the dimensionless trap frequency $1$. Nevertheless, it is remarkable that all of these states have their stable intervals, which are also summarized in Table~\ref{para1}. To check the dynamical stability of a selected state, we add as much as $1\%$ random noise in norm to the numerically exact solution but the final state is renormalized to conserve the norm, and the stability is dynamically confirmed upto $t=1000$.

The \state{0n} states closely resemble the dark-anti-dark waves reported in the literature, but both the dark and anti-dark fields have somewhat peculiar shapes. Particularly, the anti-dark solitons are not prominent, they instead take kink-like structures. In addition, the dark solitons have very unconventional equilibrium positions, e.g., the two dark solitons of \state{02} sit at the edge of the condensate rather than around the trap center and the many dark solitons do not form a regular lattice structure \cite{Dimitri:DS}. This naturally raises the question whether they are genuinely dark-anti-dark waves merely in different parameter regimes, or alternatively they are distinct states in nature. 

If they are dark-anti-dark waves, the distorted shape is most likely caused by the strong intercomponent interaction. Note that previous works studied these states in the miscible regime $g_{12}^2-g_{11}g_{22}<0$ \cite{engels16,engels20} rather than in the Manakov limit. To this end, we further continue the states and lower the intercomponent interaction $g_{12}=g_{21}$. Indeed, we find that the states gradually restore their more ``standard'' shapes as the $g_{12}$ is decreased. Interestingly, one does not need to lower $g_{12}$ very much to achieve this, some typical states at different $g_{12}$ values are shown in the bottom panels of Fig.~\ref{S01}. The state \state{010} at $g_{12} \approx 0.97$ forms a mini-lattice of dark-anti-dark solitons. These results suggest that the dispersion coefficients can have a nontrivial impact on the existence region even in the TF regime, as the dark-anti-dark waves in our setting persist all the way to the Manakov limit. In fact, the two fields appear to be in the miscible regime herein as they overlap significantly in the harmonic trap, presumably due to their different dispersion coefficients. The \state{0n} states are therefore genuinely dark-anti-dark waves.

It is interesting that the dark-anti-dark states and the in-phase dark-bright states are parametrically connected. The existence region of the \state{01} DAD soliton is approximately a reflection of that of \state{10} DB soliton about $\mu_1=\mu_2$ due to the swap of the quantum numbers, but it is slightly deformed due to our engineering of the linear limit. For a given \state{01} configuration, if we lower $\mu_1$ or alternatively increase $\mu_2$, the DAD state gradually crosses over to the DB state. Similarly, if we reduce the ``bright'' mass of the \state{02} state, this state crosses over to the two in-phase DB state \state{20} \cite{Wang:DD}. Next, the state \state{03} becomes the three in-phase DB state \state{30} \cite{Wang:DD}, and so on. 
In addition, there appears to be no sharp boundary between the DAD states and the in phase DB states, also in line with the findings of \cite{engels20}. 
However, this by no means suggests that these two series of states are of the same nature merely because they are connected by a numerical continuation.
For example, the state \state{02} is largely unstable, but the counterpart state \state{20} appears to be very robust \cite{Wang:DD}.

\subsection{Dark-multi-dark waves, and more states}

\begin{figure*}[htb]
\subfigure[]{\includegraphics[width=0.245\textwidth]{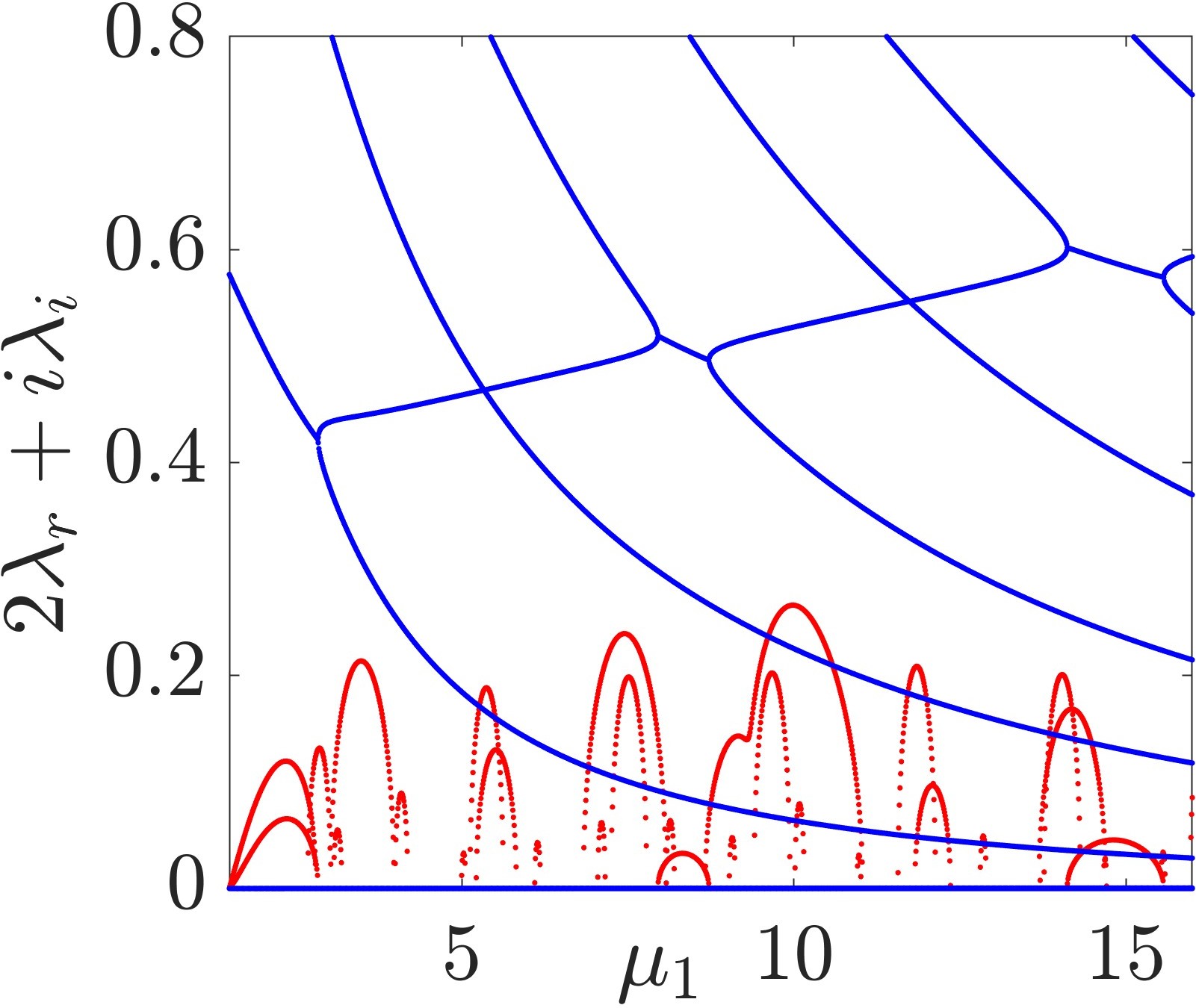}}
\subfigure[]{\includegraphics[width=0.245\textwidth]{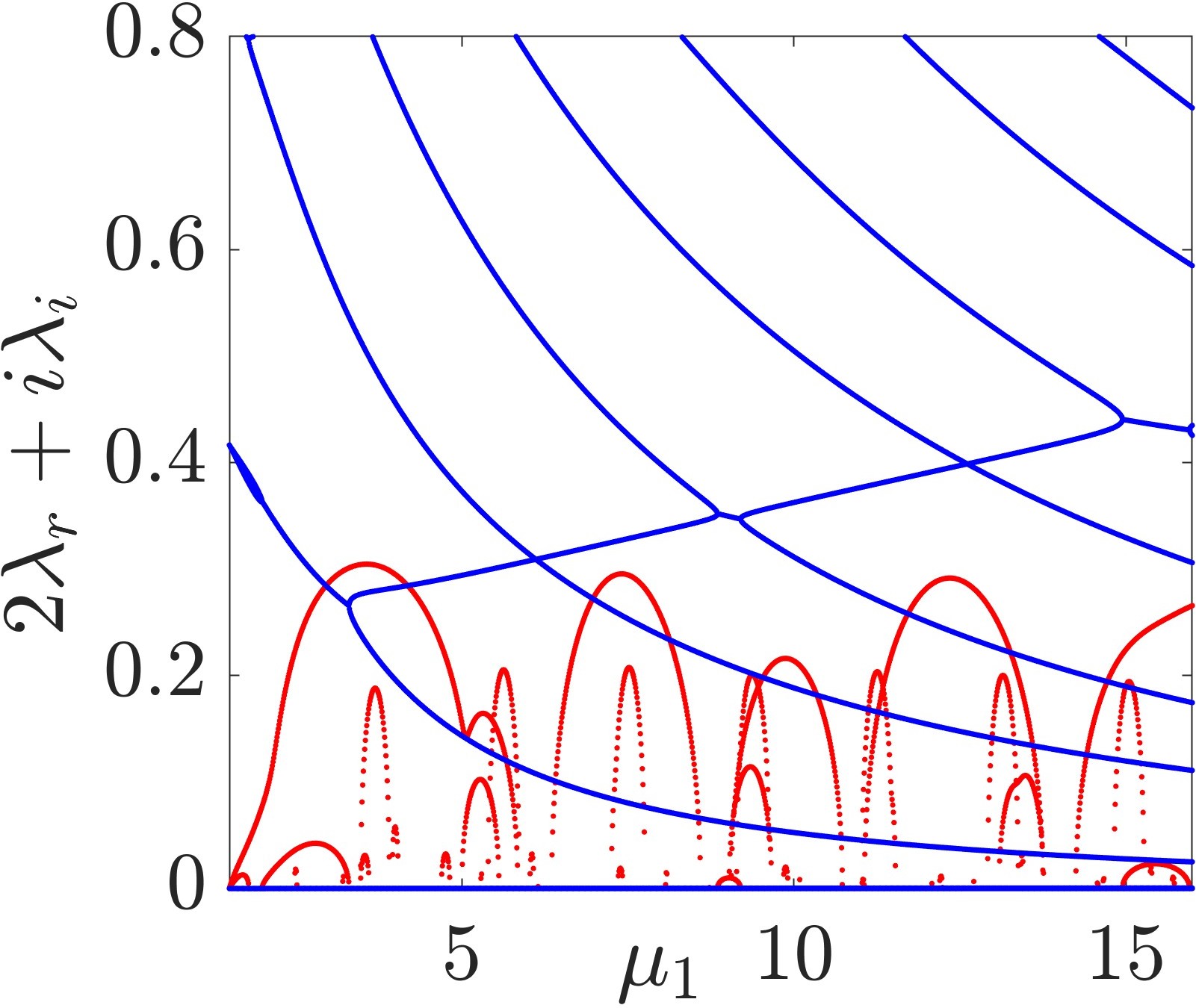}}
\subfigure[]{\includegraphics[width=0.245\textwidth]{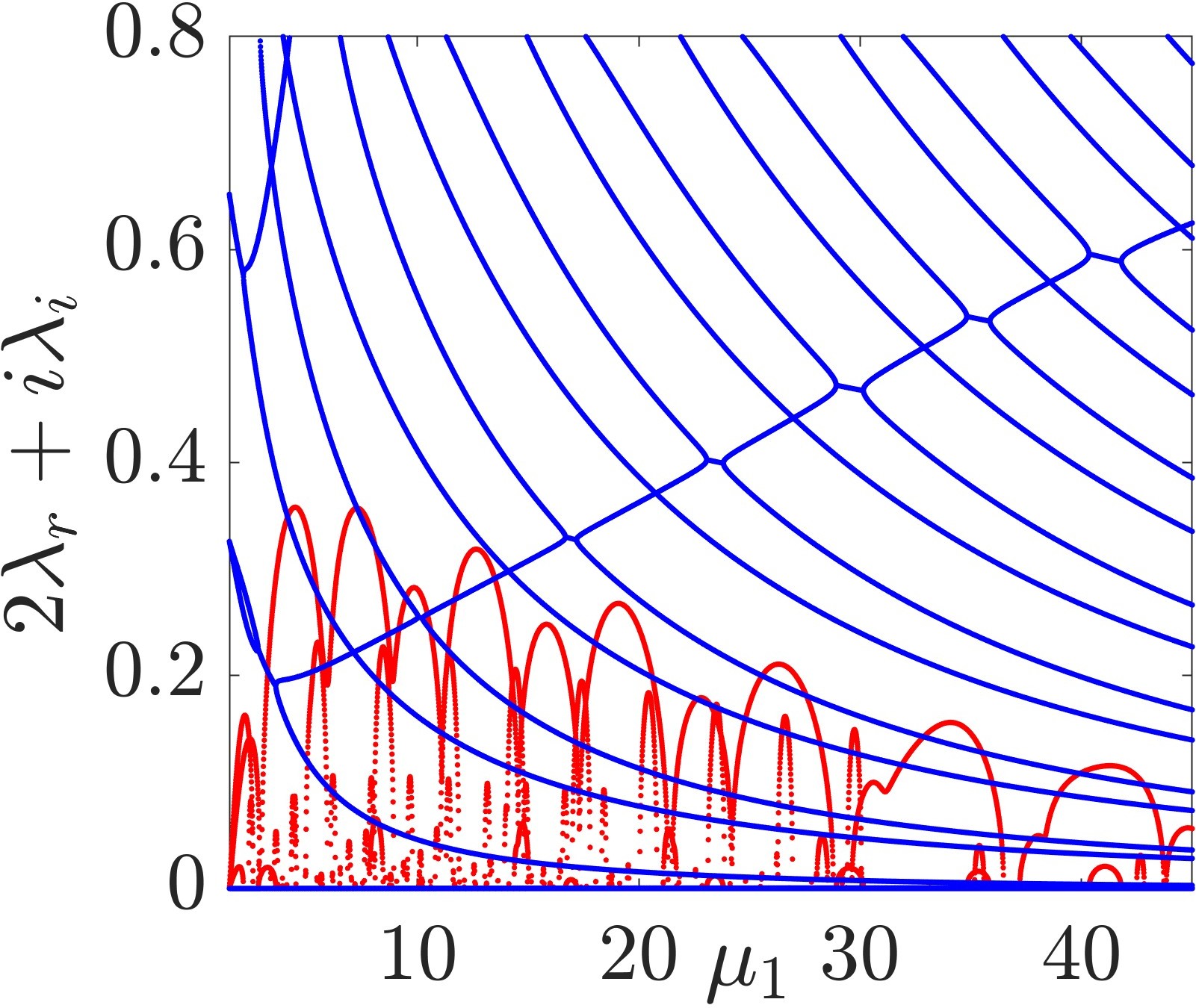}}
\subfigure[]{\includegraphics[width=0.245\textwidth]{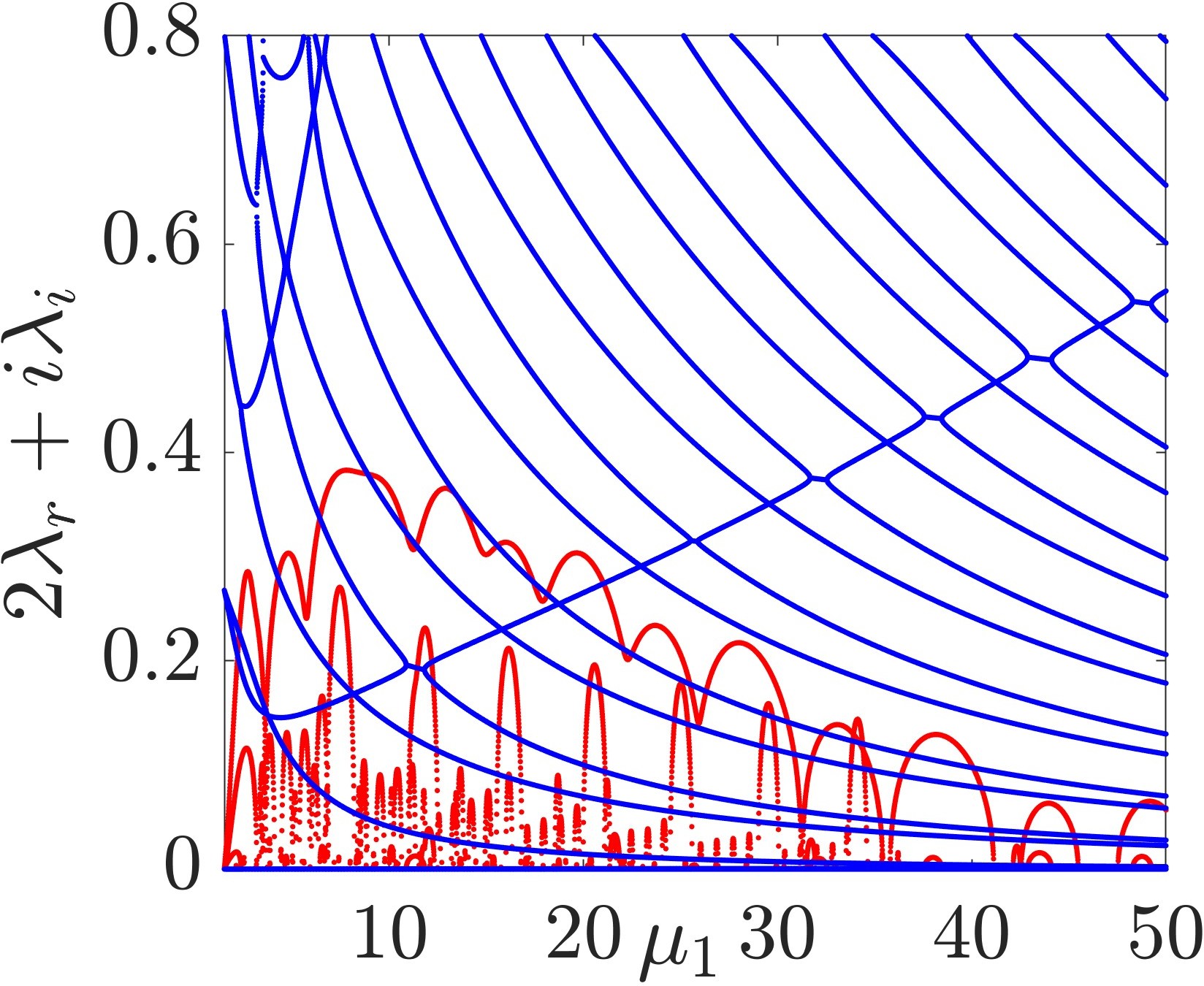}}
\subfigure[]{\includegraphics[width=\textwidth]{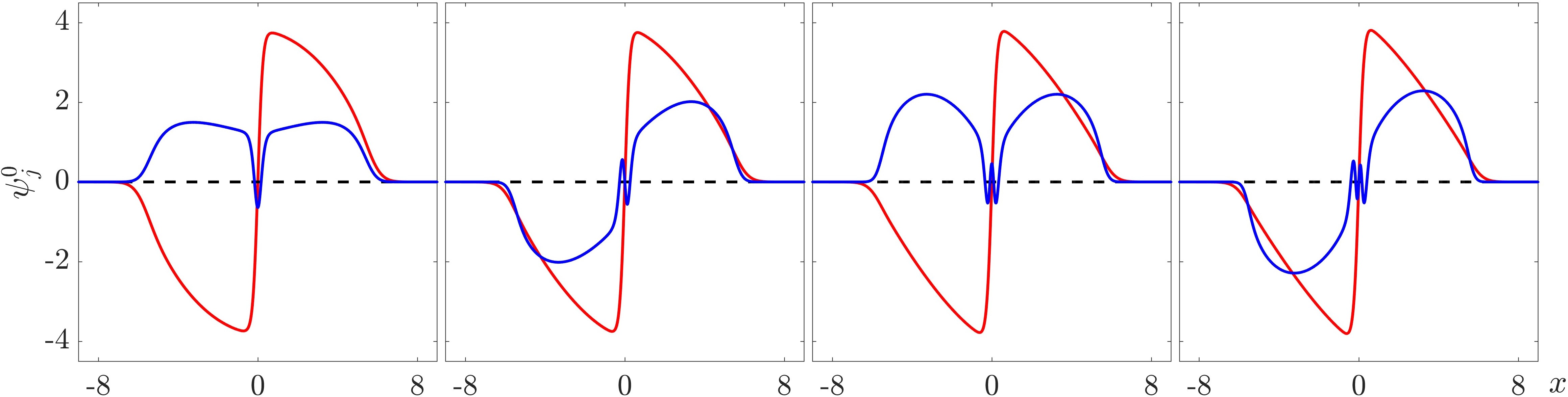}}
\caption{
Similar to Fig.~\ref{S01} but for the low-lying \state{1n} states of $n=2, 3, 4$, and $5$. Here, the dark soliton in the first component traps $2, 3, 4, 5$ dark solitons of the second component, respectively. The top panels show the BdG spectra, and typical wave configurations at $\mu_1=16$ are depicted below. The real part of the eigenvalues is enlarged by a factor of $2$ for clarity.
}
\label{S12}
\end{figure*}

The \state{1n} states for $n=2, 3, 4, 5$ and their BdG spectra are shown in Fig.~\ref{S12}. It is interesting that the dark solitons of the second component are all trapped inside the dark soliton structure of the first component, forming dark-two-dark, dark-three-dark, dark-four-dark, dark-five-dark states, respectively. This is our motivation to name these states as dark-multi-dark waves. The two fields again have significant overlapping backgrounds. The growth rate is typically much larger than that of the dark-anti-dark waves. Nevertheless, there again exist stable intervals for each of these solitary waves, as summarized in Table~\ref{para1}.

The dark solitons localization is a new feature distinct from the regular \state{n1}, $n>1$ states in \cite{Wang:DD}.
Here, we compare the structures of \state{10,1} and \state{1,10} (not shown but the trend is evident) to more clearly demonstrate the difference. In the former, the $10$ dark solitons are embedded in the background condensate, they are not localized. In the latter, the $10$ dark solitons are localized and trapped in the dark soliton of the first component.
Next, if we keep increasing the mass of the second component of \state{10,1}, the $10$ dark solitons remain very delocalized, i.e., they span the full system size and are trapped by the harmonic trap. Interestingly, the state \state{10,1} persists upon either increasing or decreasing $\mu_2$ until reaching the existence boundaries without transforming into the \state{1,10} state. Contrary to the blurred boundary of \state{10} and \state{01}, the states \state{10,1} and \state{1,10} appear to be more different in nature, at least upon changing the chemical potentials. The difference is clearly related to the engineering of the dispersion coefficients. The dark solitons localization can be appreciated from the linear limit as the spatial profile of the second wave is suppressed when $\kappa_2$ is sufficiently small, and from the TF regime as the dark solitons are attracted to the effective density potential of the dark soliton in the first component.

\begin{figure*}[htb]
\subfigure[]{\includegraphics[width=0.245\textwidth]{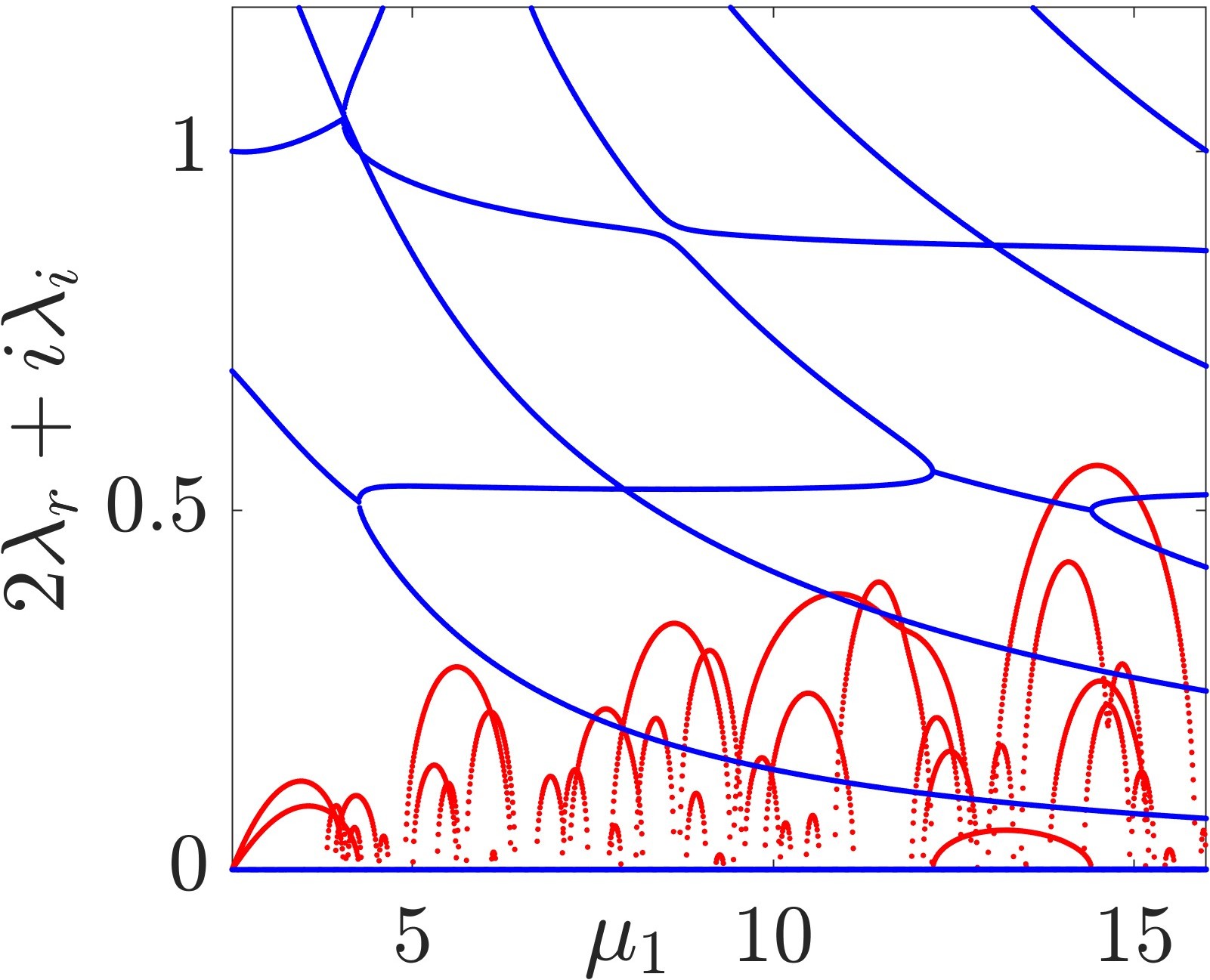}}
\subfigure[]{\includegraphics[width=0.245\textwidth]{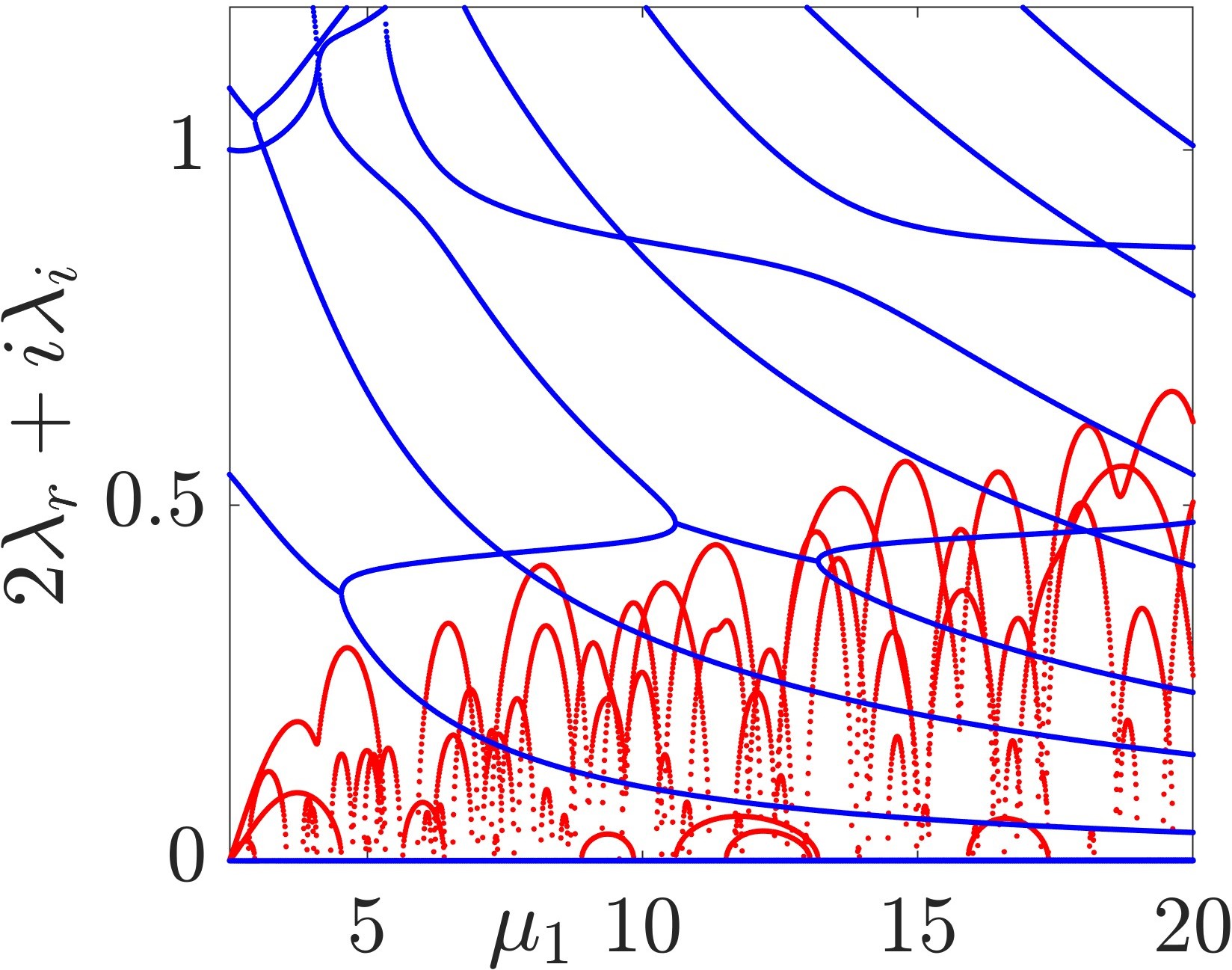}}
\subfigure[]{\includegraphics[width=0.245\textwidth]{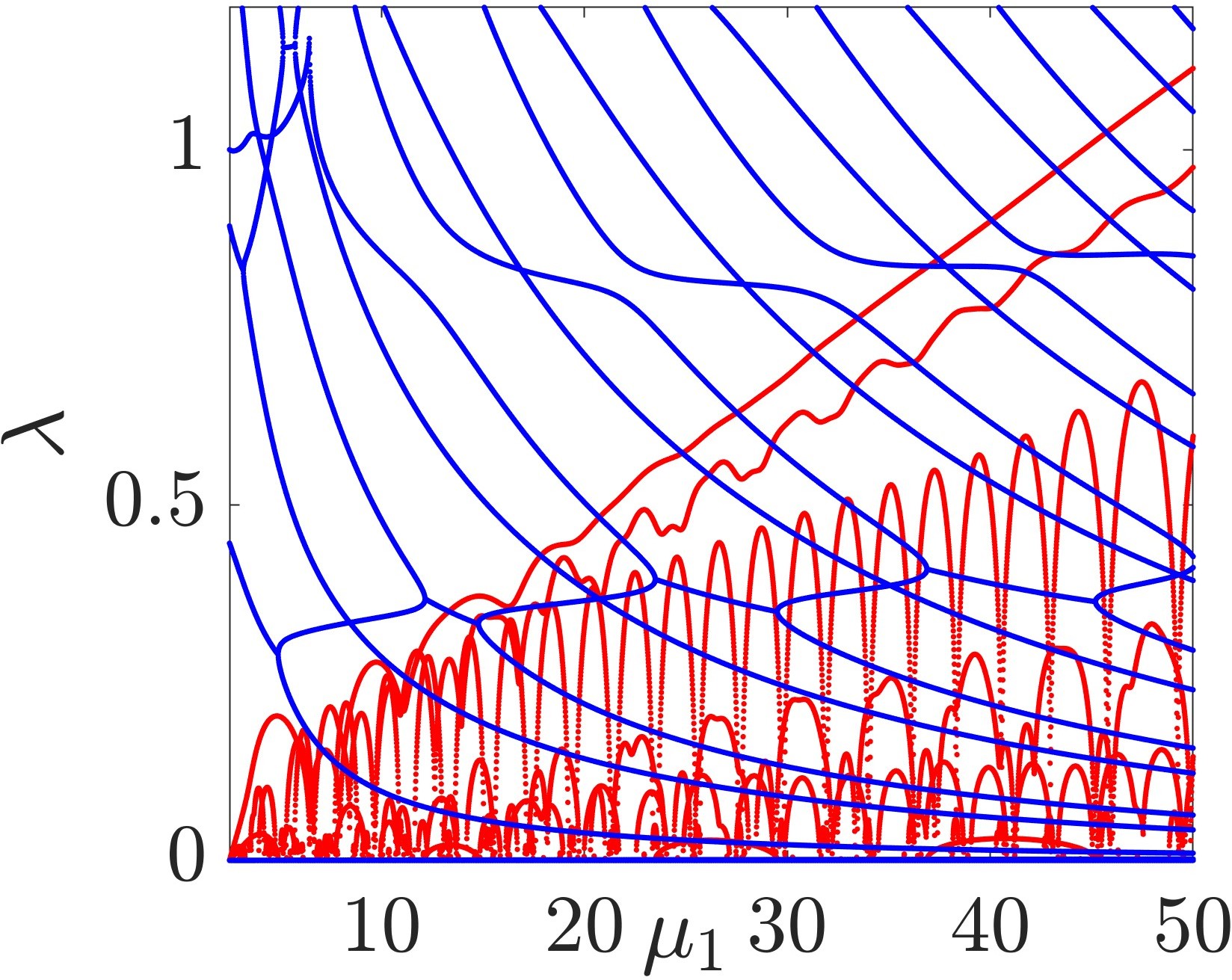}}
\subfigure[]{\includegraphics[width=0.245\textwidth]{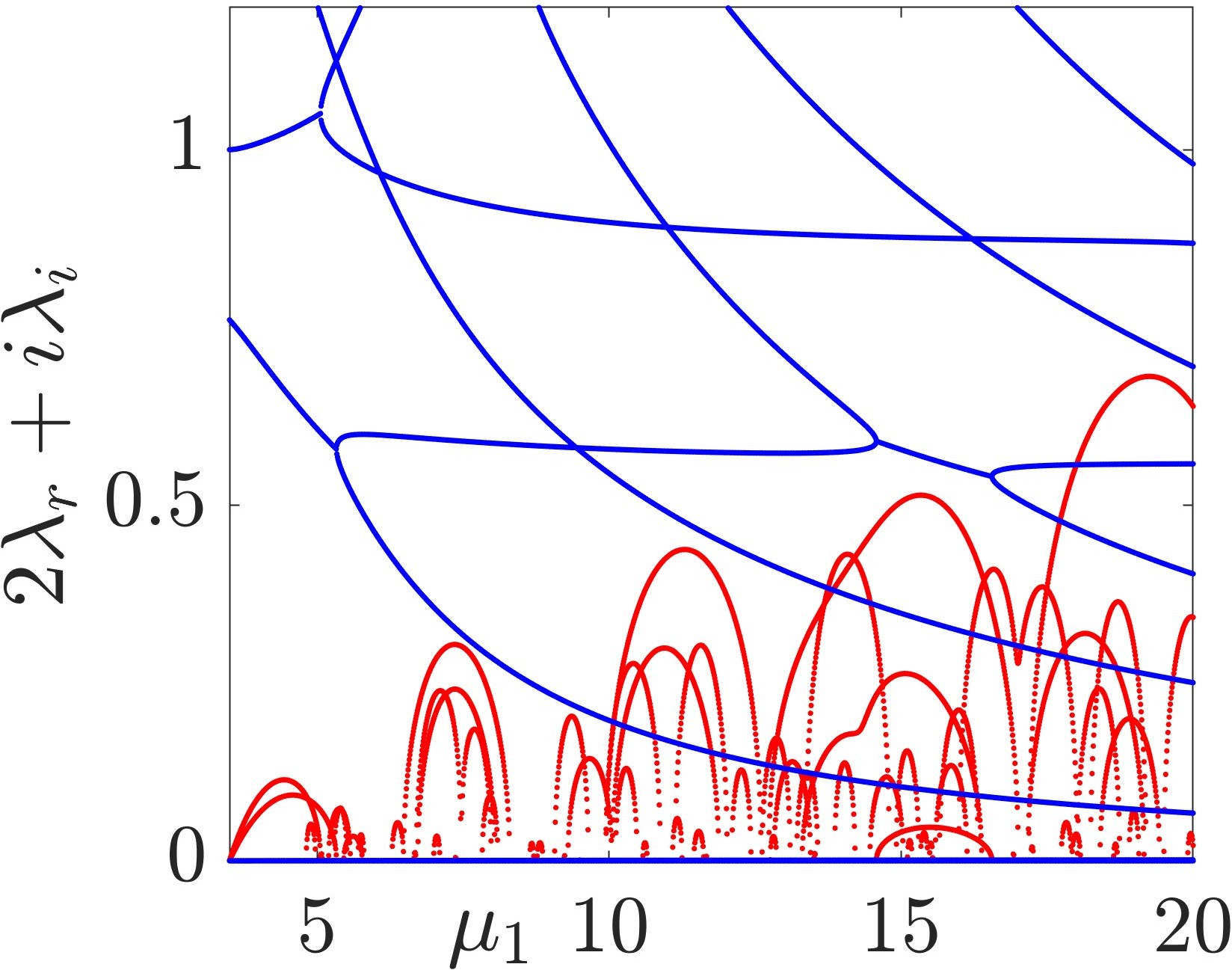}}
\subfigure[]{\includegraphics[width=0.245\textwidth]{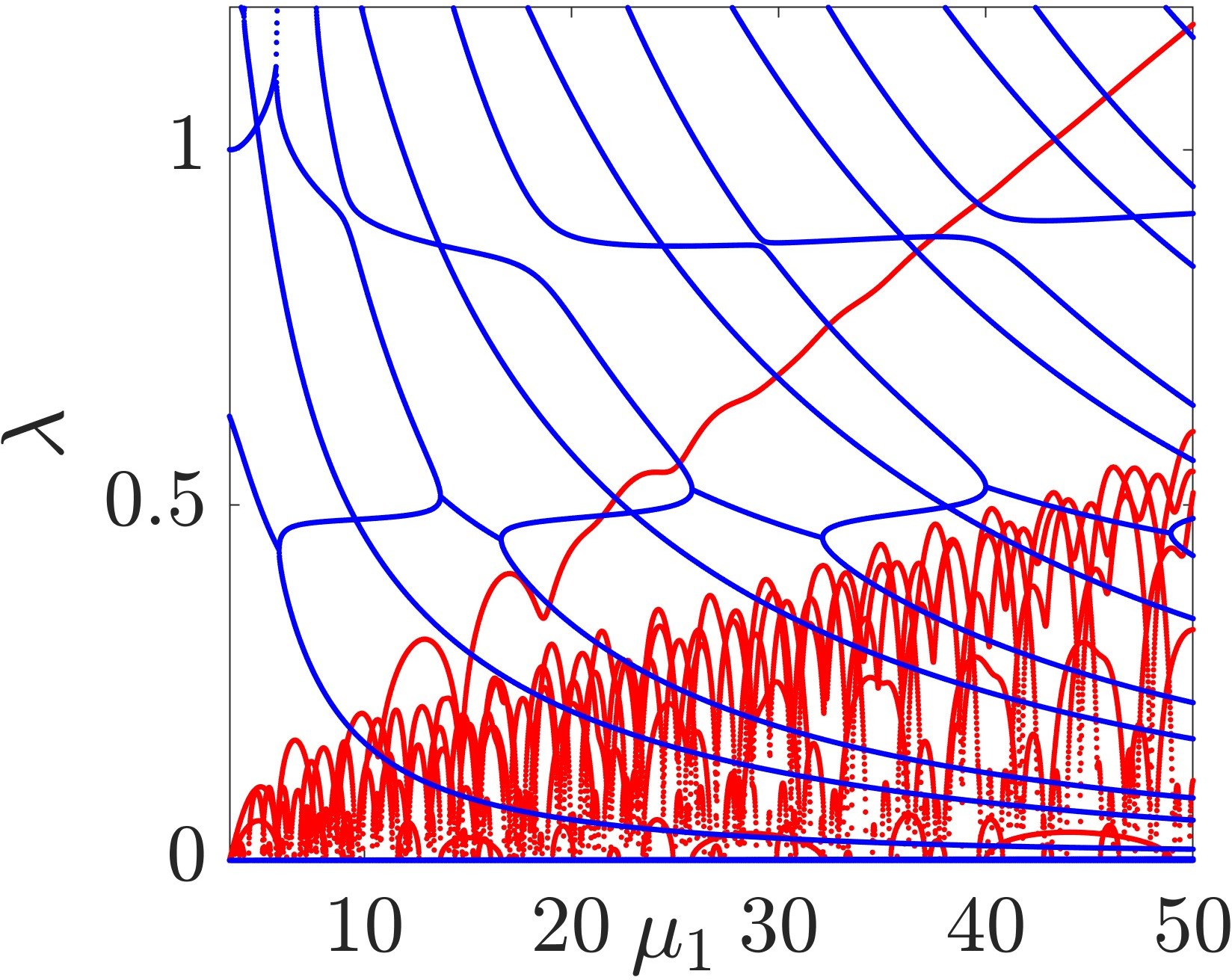}}
\subfigure[]{\includegraphics[width=0.245\textwidth]{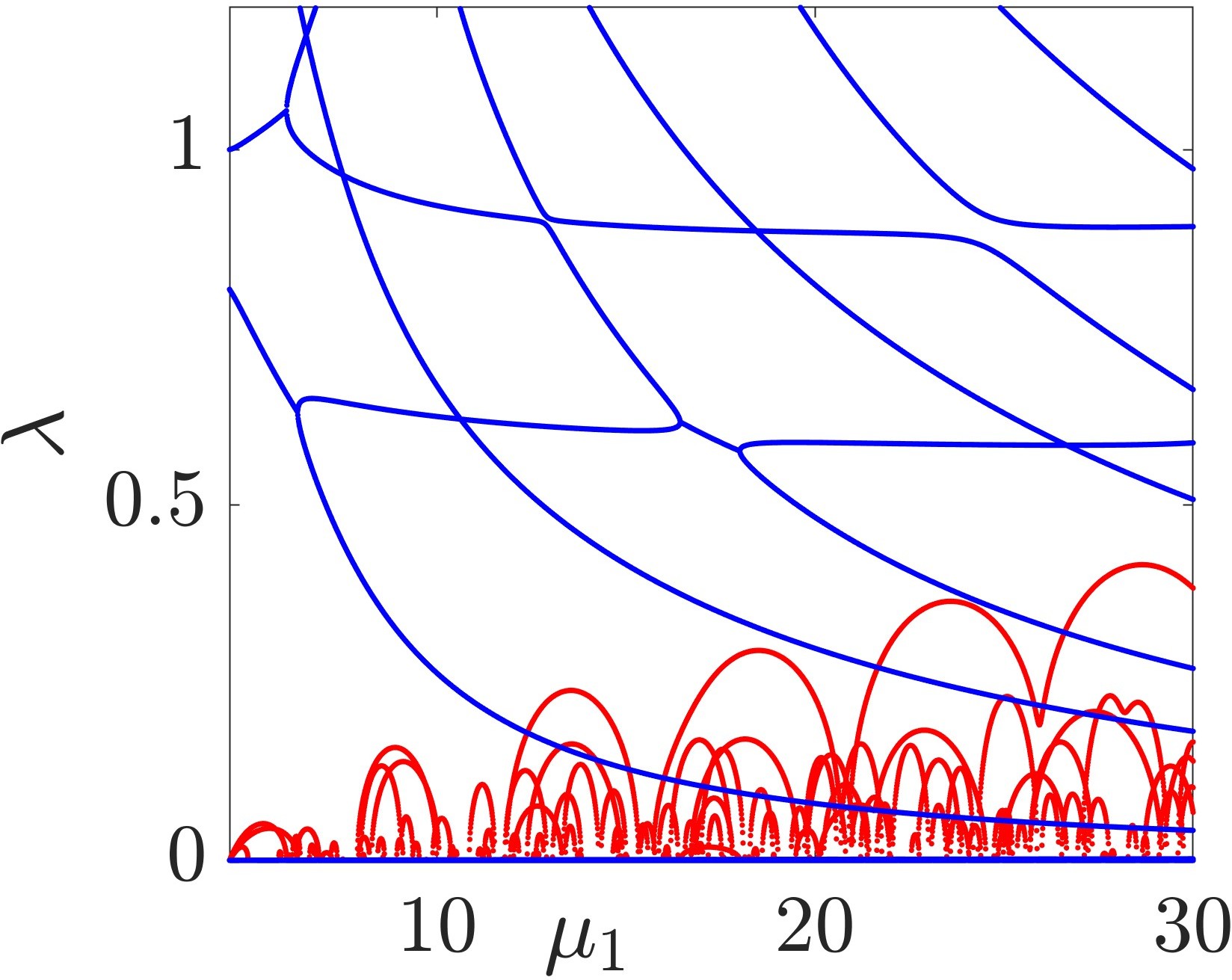}}
\subfigure[]{\includegraphics[width=0.245\textwidth]{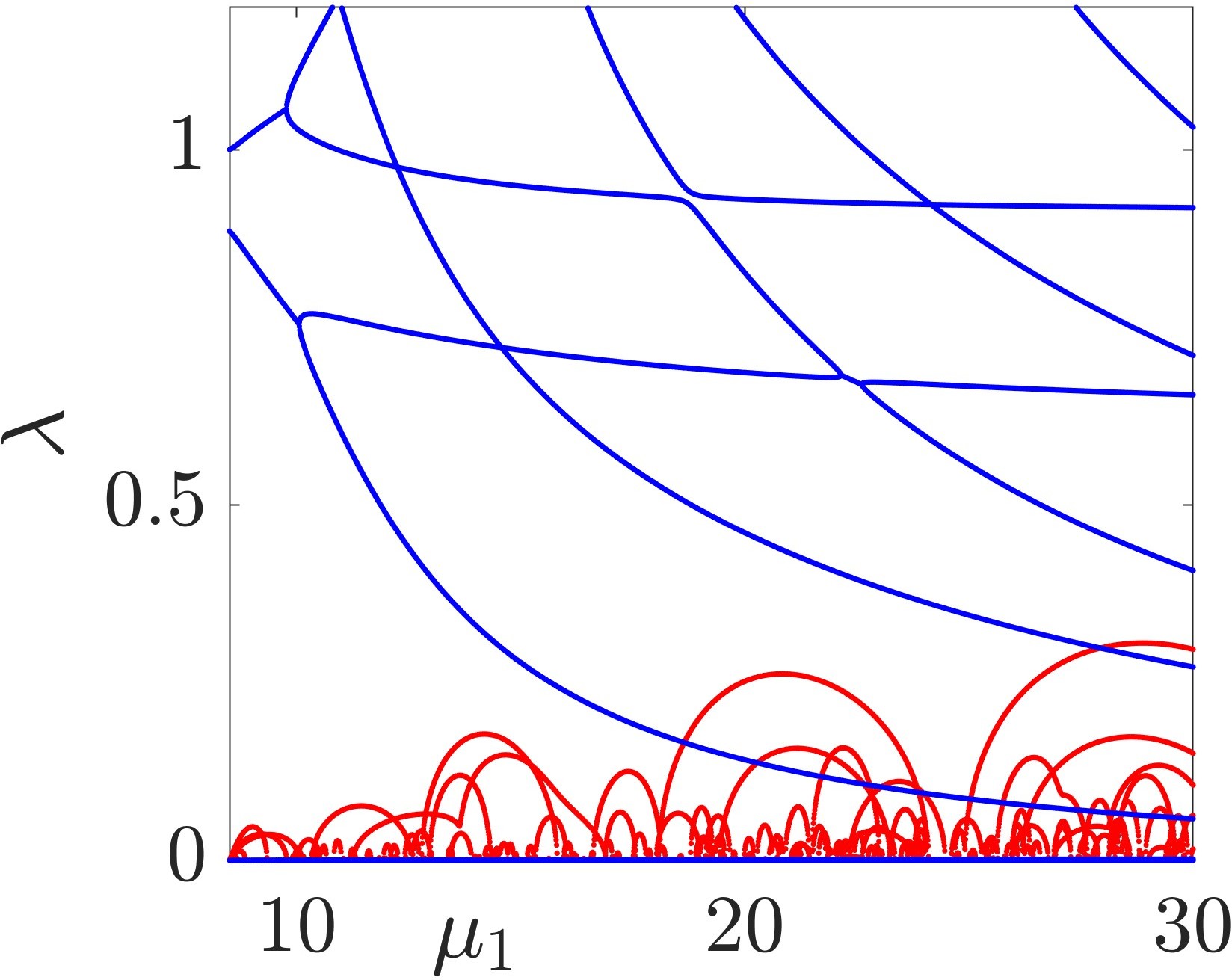}}
\subfigure[]{\includegraphics[width=0.245\textwidth]{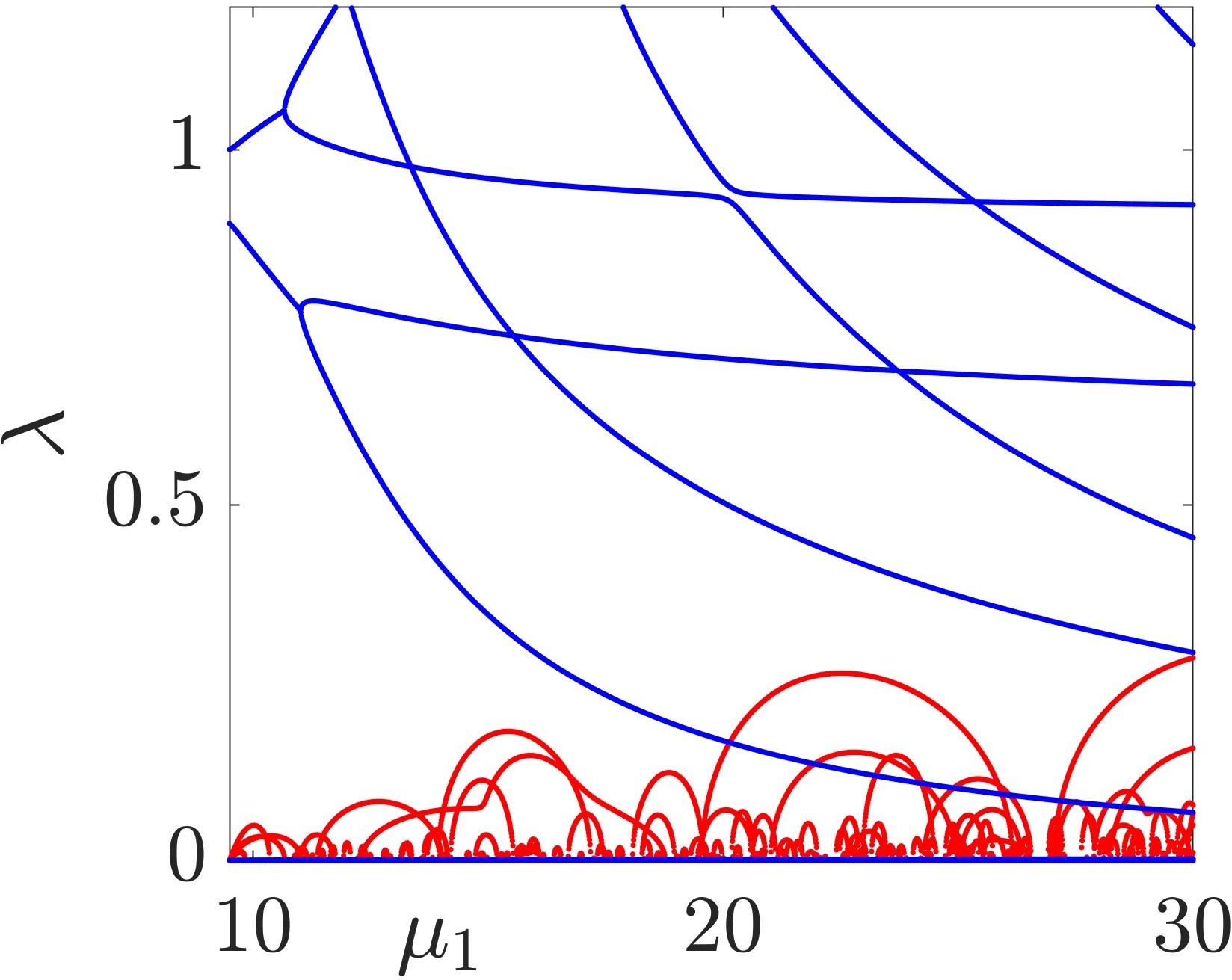}}
\subfigure[]{\includegraphics[width=\textwidth]{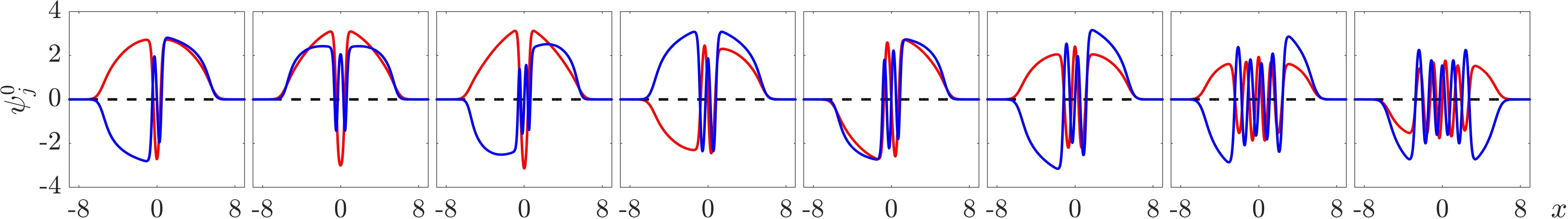}}
\caption{
Similar to Fig.~\ref{S12} but for some more excited states, these states are the \state{2n} states of $n=3, 4, 5$, the \state{3n} states of $n=4, 5$, and a few even more excited states \state{45}, \state{89}, and \state{9,10}, respectively. As before, typical wave configurations at $\mu_1=16$ are depicted.
}
\label{S23}
\end{figure*}

The construction is systematic in nature and is not limited to the \state{0n} and \state{1n} states, i.e., there are \state{2n} states, \state{3n} states, and so on. It is clearly beyond the scope of this work to examine each series in detail. Here, we illustrate that these states are indeed available and present some prototypical low-lying states and a few highly excited states. Some typical configurations and their BdG spectra are depicted in Fig.~\ref{S23}. While we calculated the BdG spectra only for these waves, we have systematically verified the existence of the waves of \state{n_1n_2} for all the cases of $0\leq n_1<n_2\leq 10$, therefore, the linear limit continuation method appears to be very robust in the setting of unequally dispersion coefficients.


Two common features are present in all of the states considered herein. First, the two fields are effectively miscible as they strongly overlap.
Second, the multiple dark solitons of the second component are approximately trapped in the dark solitons region of the first component, except obviously the dark-anti-dark waves. It is interesting that hierarchical structures can emerge, the \state{24} state is consist of two balanced \state{12} structures around the trap center. This pattern formation trend continues, the \state{36}, \state{48}, \state{5,10} states (not shown) in turn are consist of three, four, five \state{12} units that are approximately equally spaced.

In general, it is impossible to decompose a complex wave into an array of more elementary structures, similar to \cite{Wang:DD}. In the state \state{23}, the two dark solitons of the first component are close to the two outside dark solitons of the second component, the central dark soliton of the second component traps the central mass of the first component. One can perhaps view this structure as approximately an array of dark-dark, dark-bright, and dark-dark solitons. In the state \state{25}, the structure is similar expect now each dark soliton in the first component traps two dark solitons, forming approximately an array of dark-two-dark, dark-bright, dark-two-dark solitons. We shall not discuss this decomposition further for the more complicated states here.

The stability deteriorates further for these more excited waves in both the number of unstable modes and the growth rates as expected, but the trend is not very monotonic in the quantum numbers. For example, it seems challenging to stabilize the \state{25} and \state{35} states, but it is relatively easier to stabilize the more excited \state{89} and \state{9,10} states. This is presumably because in the former, the quantum numbers are large and also the dispersion coefficients differ significantly. Both factors may make a state less easier to stabilize. Nevertheless, most states can be fully stabilized as summarized in Table~\ref{para1}.


\begin{figure*}[htb]
\subfigure[]{\includegraphics[width=0.495\textwidth]{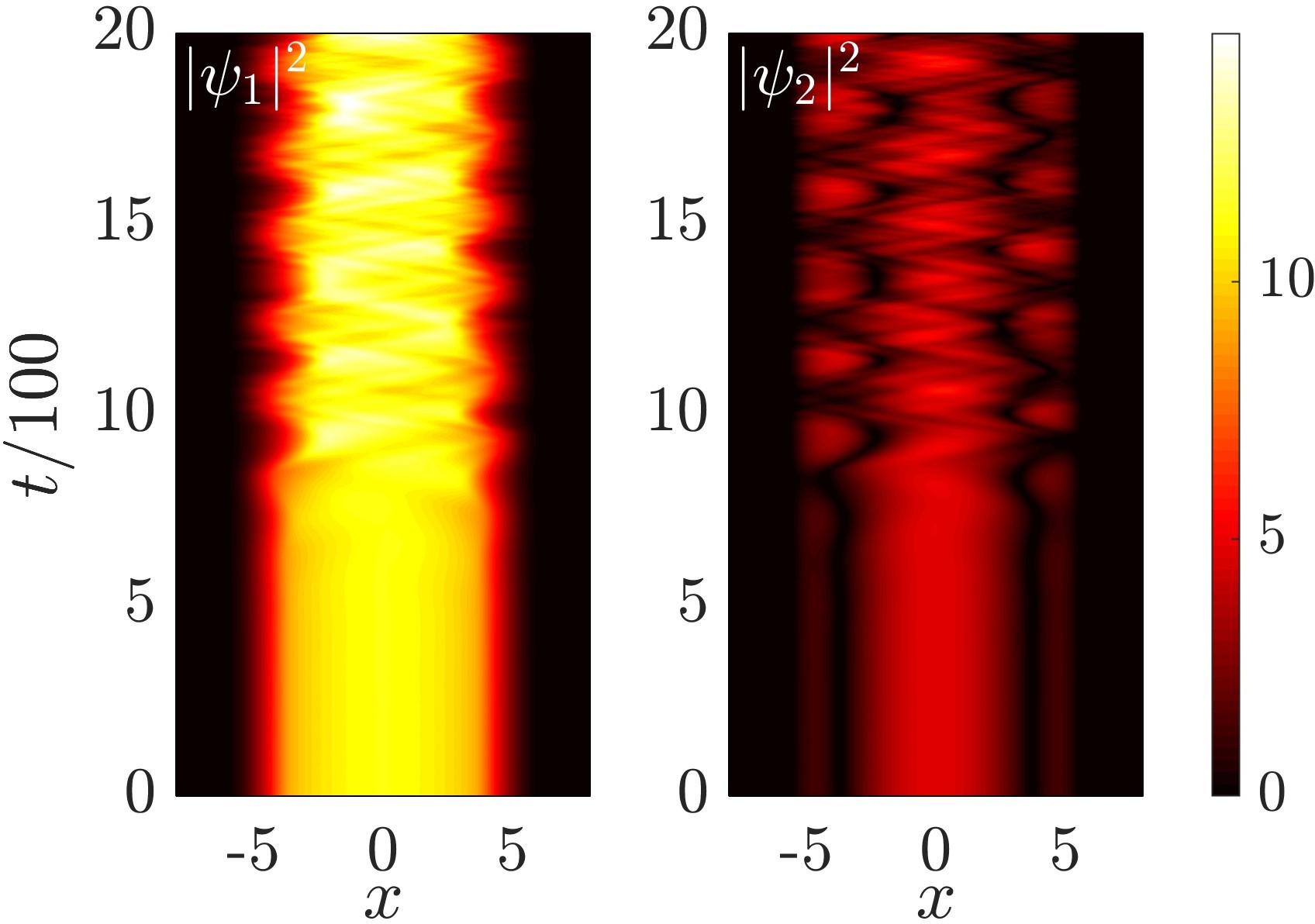}\label{Dy1}} 
\subfigure[]{\includegraphics[width=0.495\textwidth]{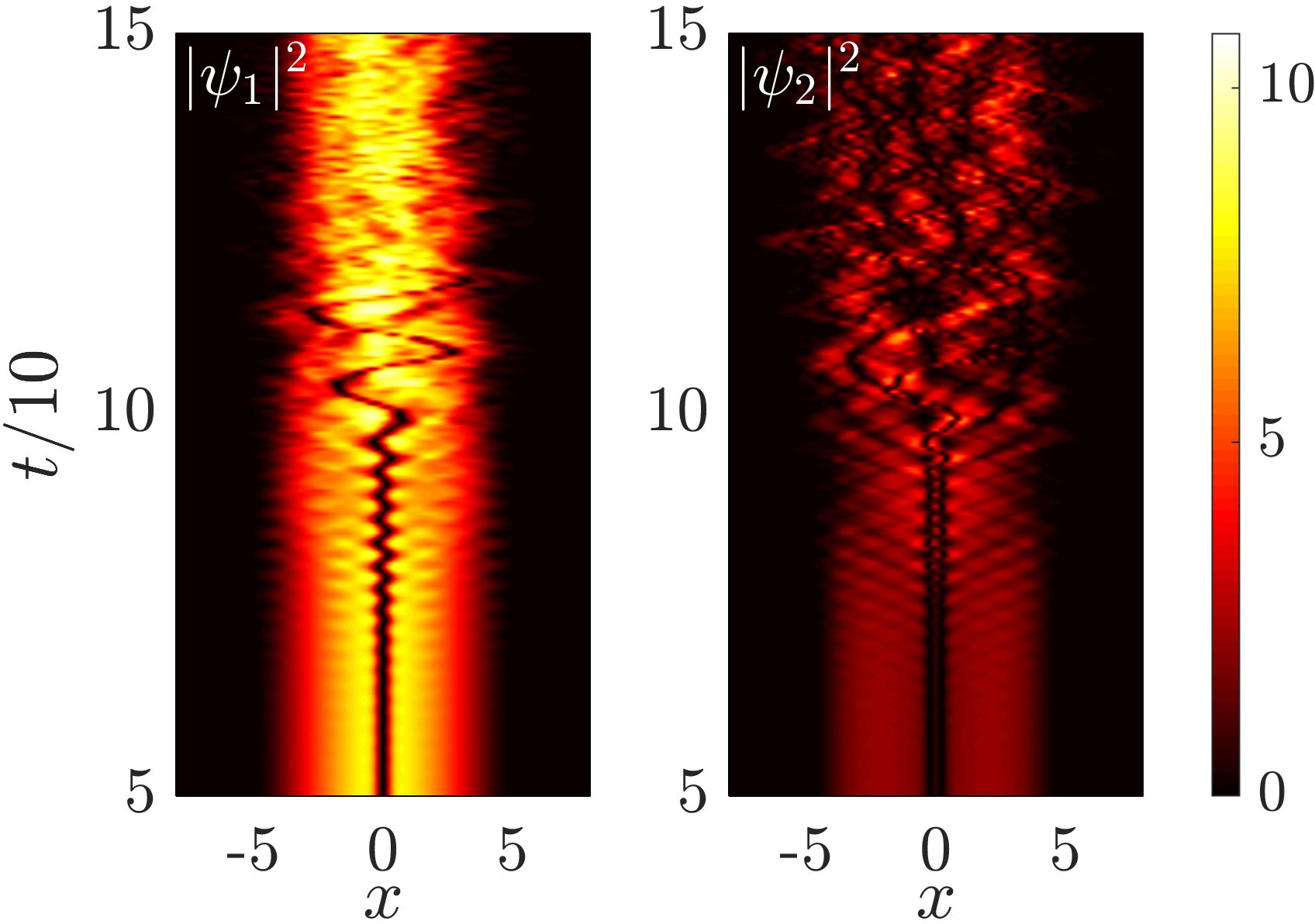}\label{Dy2}} 
\caption{Dynamical instability of the most unstable mode of states \state{02} (a) and \state{12} (b) at $\mu_1=16$ and $\mu_1=10.3$, respectively. In the \state{02} instability, the two dark-anti-dark in-phase oscillation is excited and becomes prominent around $t=700$. In the \state{12} instability, the two dark solitons of the second component are in phase, and they are out-of-phase with respect to the central dark soliton of the first component. Both dynamics lead to complex dark-anti-dark oscillation patterns, see the text for more details.
}
\label{Dy}
\end{figure*}

Finally, we illustrate two unstable scenarios of dynamics. It is clearly beyond the scope of this work to systematically investigate all the dynamical instabilities of these large arrays of states. Here, we only present two simple show-of-principle dynamics to illustrate the rich behaviours of these waves. To this end, we examine the most unstable mode of \state{02} and \state{12} states at $\mu_1=16$ and $\mu_1=10.3$, respectively. Here, we reduce the noise level to $0.1\%$ to probe the most unstable mode more clearly.

In the \state{02} unstable dynamics shown in Fig.~\ref{Dy1}, the in-phase oscillation of the dark-anti-dark solitons is excited. The oscillation becomes prominent at around $t=700$, and the two dark solitons quickly run out of  synchronization in their velocities presumably because of the strongly overlapping central clouds, leading to complex multiple dark-anti-dark solitons dynamics and density excitations in the condensates.

The instability onset of the \state{12} state is much faster, in line with its much larger growth rate, as illustrated in Fig.~\ref{Dy2}. In this dynamics, the two dark solitons of the second component oscillate in phase, but the two together are out of phase with respect to the central dark soliton of the first component. The oscillation amplitude gradually grows, and at $t\approx 70$, the three dark soliton velocities lost their synchronization. The central dark soliton eventually escapes from the first component, while more dark solitons are induced into the second component from the edge, forming interestingly also a complex mixture of interacting dark-anti-dark solitons.

\section{Conclusions and Future Challenges}
\label{conclusion}

In this work, we presented a systematic construction of vector solitary waves from their linear limits in the setting of unequal dispersion coefficients. The method is robust, and yields new series of solitary waves compared with the regular ones obtained earlier \cite{Wang:DD}. Particularly, the lowest-lying series corresponds to the well-known dark-anti-dark waves, the next series yields the dark-multi-dark waves, and more complex wave patterns can also be constructed in a systematic manner. While these states typically have less favourable stability properties, most states can be fully stabilized in suitable parameter regimes. Finally, two proof-of-principle dynamics are illustrated, showing the rich behaviours of the waves. This work complements the earlier one \cite{Wang:DD}, and significantly expands the type and number of solitary waves available from the linear limit continuation method, and therefore provides a coherent framework to understand a diverse set of solitary waves, most notably the dark-bright, dark-dark, dark-anti-dark, and dark-multi-dark structures.

This work can be generalized in multiple directions in the next. First, it is interesting to apply the method to the one-dimension setting with three or more components \cite{Wang:MDDD} following a similar setup of ordering the linear eigenenergies with different dispersion coefficients. Our preliminary result shows that the regular low-lying state \state{210} of three components \cite{Wang:MDDD} can be generalized to states of all possible permutations of the quantum numbers, i.e., states \state{012}, \state{021}, \state{102}, \state{120}, \state{201}, and \state{210}. It is clear that these waves are not identical, however, future work should examine in detail whether it is possible to distinguish these states in a straightforward manner because these waveforms are quite complex and cannot be readily decomposed into a collection of elementary structures such as dark-dark-bright and dark-bright-bright solitons \cite{Katsimiga_2021,Lichen:DBB}. If we assume these waves are distinct, the number of solitary waves from the linear limit continuation method grows by a factor of $n!$ in the $n$-component system.

It is particularly interesting to extend the work to the much more versatile higher dimensions. However, as a first step, the method should be systematically developed in the single component system due to the presence of degenerate states as discussed in \cite{Wang:MDDD}. Then, two linear states in a two-component system can be coupled either as in \cite{Wang:DD} or as here with different dispersion coefficients. If one is only interested in a particular class of waves, it is likely possible to directly study these waves. For example, one can directly study three-dimensional dark-anti-dark waves. Two highly interesting structures would be the vortex-line-anti-dark and vortex-ring-anti-dark solitary waves. In two dimensions, a vortex may trap a vortex cluster similar to the dark-multi-dark waves herein. Research efforts along some of these directions are currently in progress
and will be reported in future publications.



\begin{acknowledgments}
We gratefully acknowledge supports from the National Science Foundation of China under Grant No. 12004268, the Fundamental Research Funds for the Central Universities, China, and the Science Speciality Program of Sichuan University under Grant No. 2020SCUNL210.
We thank the Emei cluster at Sichuan
University for providing HPC resources.
\end{acknowledgments}

\bibliography{Refs}

\end{document}